\documentclass[a4paper,11pt]{article}
\pdfoutput=1
\usepackage{jheppub}
\usepackage{amsmath,shuffle,comment,amsfonts}
\usepackage{empheq}
\usepackage{graphicx} 
\usepackage{xcolor}
\usepackage{booktabs}
\usepackage{tikz,diagbox}
\usetikzlibrary{calc}
\usepackage{setspace}
\def\ma{\mathcal}
\def\ie{\begin{equation}\begin{aligned}}
\def\fe{\end{aligned}\end{equation}}
\usepackage{enumitem}
\usetikzlibrary{decorations.markings, arrows.meta, calc, decorations.pathmorphing,angles, quotes, arrows.meta}

\usepackage{hyperref}
\hypersetup{
  linktocpage,
  colorlinks  = true, 
  urlcolor    = magenta, 
  linkcolor   = magenta, 
  citecolor   = purple 
}

\title{All-multiplicity monodromy and KLT relations for AdS string integrals}

\author[1]{Maria Nocchi,}
\author[2]{Rodrigo Schmidt Pitombo,}
\author[3]{Aurélie Strömholm Sangaré,}
\author[4]{and Yi-Xiao Tao}

\affiliation[1]{Simons Center for Geometry and Physics, Stony Brook, NY 11794, USA}
\affiliation[2]{ICTP South American Institute for Fundamental Research,
  Instituto de Física Teórica,\\ Universidade Estadual Paulista (UNESP), São Paulo 01140-070, SP, Brazil}
\affiliation[3]{Mathematical Institute, University of Oxford,\\
  Andrew Wiles Building, Woodstock Road, Oxford OX2 6GG, United Kingdom}
\affiliation[4]{Department of Mathematical Sciences, Tsinghua University, Beijing 100084, China}

\emailAdd{maria.nocchi@stonybrook.edu}
\emailAdd{rs.pitombo@unesp.br}
\emailAdd{aurelie.sangare@maths.ox.ac.uk}
\emailAdd{taoyx21@mails.tsinghua.edu.cn}

\abstract{
We propose and study all-multiplicity building blocks for tree-level string amplitudes in AdS. These are worldsheet integrals obtained by dressing the corresponding flat-space disc and sphere integrals with multivariable multiple polylogarithms and their single-valued analogues, respectively. We derive monodromy relations for the open-string building blocks and a KLT factorisation for their closed-string counterparts. This extends the non-commutative AdS uplift of lower-point flat-space structures to general \(n\)-point kinematics.
}

\begin{document}
\setlength{\abovedisplayskip}{10pt plus 2pt minus 2pt}
\setlength{\belowdisplayskip}{10pt plus 2pt minus 2pt}
\setlength{\abovedisplayshortskip}{10pt plus 2pt minus 2pt}
\setlength{\belowdisplayshortskip}{10pt plus 2pt minus 2pt}

\maketitle
\section{Motivation and summary}\label{sec:Introduction}

A central lesson of perturbative string theory is that scattering amplitudes are computed from worldsheet correlation functions of vertex operators, integrated over moduli spaces of punctured Riemann surfaces, rather than assembled from local point-particle vertices. At tree level, the relevant geometries are the disc for open strings and the sphere for closed strings. Nevertheless, string amplitudes exhibit an organisation reminiscent of field-theory amplitudes. Many of these structures have a natural worldsheet origin: cyclicity, factorisation, monodromy~\cite{Plahte:1970wy} and Kawai--Lewellen--Tye (KLT) relations~\cite{Kawai:1985xq} are encoded in, and can often be derived from, the geometry of string moduli spaces~\cite{Stieberger:2009hq,Broedel:2013tta}.

For open strings on the disc, the gauge degrees of freedom carried by the string endpoints are encoded in Chan--Paton factors, represented by matrices acting on gauge indices. Separating these factors from the kinematic part of the amplitude leads to a decomposition into colour-ordered partial amplitudes, each associated with a fixed cyclic ordering of the vertex operators on the boundary. Let $z_i$ denote the position of the $i$-th boundary vertex operator, or puncture. After gauge fixing to \((z_1,z_{n-1},z_n)=(0,1,\infty)\), an ordered \(n\)-point tree-level contribution takes the schematic form
\begin{equation}
  A_n^{\rm open}(1,2,\ldots,n)
  \sim
  \int_{0<z_2<\ldots<z_{n-2}<1} \,
  \prod_{i=2}^{n-2} dz_i\,
  \prod_{i<j} |z_{ij}|^{s_{ij}}\,
  \mathcal F_n(z) \,,
  \label{eq:open-string-gauge-fixed}
\end{equation}
where the Koba--Nielsen factor \(\prod_{i<j}|z_{ij}|^{s_{ij}}\) captures the universal dependence on the external momenta, while \(\mathcal F_n(z)\) encodes the remaining information about the external states and the particular string theory under consideration.

The geometry of the ordered boundary imposes several constraints on the partial amplitudes, including cyclicity, reflection and monodromy. The latter arises from analytically continuing punctures through the ordered integration domain. A representative identity takes the schematic form
\begin{equation}
  A_n^{\rm open}(1,2,\ldots,n)
  +
  \sum_{j=2}^{n-1}
  e^{i\pi(s_{12}+s_{13}+\ldots+s_{1j})}\,
  A_n^{\rm open}(2,3,\ldots,j,1,j{+}1,\ldots,n)
  =
  0 \,.
  \label{eq:schematic-monodromy-intro}
\end{equation}
The phases arise from the analytic continuation of the Koba--Nielsen factor. Taken together, these properties reduce the independent partial amplitudes to a basis of size \((n-3)!\). In the field-theory limit, they reproduce the Kleiss--Kuijf relations~\cite{Kleiss:1988ne} and the Bern--Carrasco--Johansson (BCJ) relations~\cite{Bern:2008qj} among colour-ordered gauge-theory amplitudes~\cite{Stieberger:2009hq,Bjerrum-Bohr:2009ulz}.

For closed strings on the sphere, the amplitude is controlled by the pairing of holomorphic and anti-holomorphic worldsheet structures, encoded in the KLT relations
\begin{equation}
  M_n^{\rm closed}
  =
  A_n^{{\rm open}\,\mathsf T}\,
  S_{\alpha'}\,
  \widetilde A_n^{\rm open}\,,
  \label{eq:klt-intro}
\end{equation}
where the open-string amplitudes are understood as vectors in a basis of colour orderings and the superscript \(\mathsf T\) denotes transposition in this space. The momentum kernel \(S_{\alpha'}\), whose all-multiplicity form was made explicit in~\cite{Bjerrum-Bohr:2010pnr}, is built from the sine factors of the Mandelstam variables dictated by open-string monodromy. Thus, already at tree level, the geometry of the worldsheet relates the factorisation of closed-string sphere integrals to the analytic continuation properties of open-string disc integrals.

This same structure makes string amplitudes a natural laboratory for special functions and periods. In quantum field theory, transcendental functions such as multiple polylogarithms typically arise from loop integrals \cite{Duhr:2014woa}. In string theory, by contrast, analogous structures already appear at tree level. The low-energy expansion of disc integrals produces multiple zeta values, while closed-string sphere integrals involve their \textit{single-valued} counterparts~\cite{Schlotterer:2012ny,Stieberger:2013wea,Stieberger:2014hba,Brown:2019wna}. In this sense, closed-string tree amplitudes can be viewed as a single-valued projection of open-string amplitudes.\footnote{At higher genus, this is enriched by elliptic and modular generalisations  (see for example~\cite{Broedel:2014vla,DHoker:2015wxz,Schlotterer:2025qjv}).}

The tree-level worldsheet representation is conceptually simple, but the moduli-space integrals encode much of the physics. Factorisation and the field-theory limit arise from degenerations of the worldsheet, while the finite-\(\alpha'\) dependence involves special functions of increasing complexity at higher multiplicity. In flat space, this dependence can be organised into universal worldsheet integrals, or \textit{building blocks}, separating the moduli-space dependence from the theory-specific kinematic data.

Open-string partial amplitudes can be expanded in a basis of Yang--Mills amplitudes multiplied by disc integrals carrying the full \(\alpha'\)-dependence~\cite{Mafra:2011nv,Mafra:2011nw}. Generalising the Euler beta function of the four-gluon Veneziano amplitude, these integrals can be decomposed into doubly ordered integrals \(Z(\tau|\rho)\), where \(\tau\) specifies the integration cycle and \(\rho\) the Parke--Taylor ordering, with $\tau,\rho \in S_n$. They are the basic objects of \(Z\)-theory~\cite{Mafra:2016mcc}. Closed-string amplitudes similarly involve doubly ordered sphere integrals.

The flat-space picture provides both a guide and a contrast for analogues of string amplitudes in AdS, where there is no conventional S-matrix for asymptotic states. The natural observables are instead boundary correlation functions of the dual conformal field theory~\cite{Maldacena:1997re,Witten:1998qj}. It is therefore non-trivial to translate flat-space amplitude structures into AdS. Significant recent progress suggests that a worldsheet-like structure continues to organise string amplitudes in AdS~\cite{Alday:2023mvu,Alday:2024yax,Wang:2025owf,Wang:2025pjo}. At four points, the tree-level low-energy expansion is governed by the same class of special numbers that appears in flat space, namely multiple zeta values. More generally, AdS open- and closed-string building blocks are organised into two towers of integrals obtained by inserting multiple polylogarithms and their single-valued counterparts into the flat-space disc and sphere integrals.\footnote{Further support for the proposed worldsheet structure comes from high-energy limits~\cite{Alday:2023pzu,Alday:2024xpq} and different AdS backgrounds~\cite{Chester:2024wnb,Chester:2024esn,Alday:2024rjs}. Checks against localisation and integrability are also discussed there.} These two towers are related by an AdS version of the KLT relations~\cite{Alday:2025bjp,Kakkad:2025klm}, and the open-string building blocks further obey AdS monodromy relations~\cite{Alday:2025cxr}.

This paper aims to extend this picture to arbitrary multiplicity. Four-point AdS building blocks already capture non-trivial stringy deformations of the beta-function structure, but higher multiplicity probes new features of moduli space and its boundary stratification. From the holographic perspective, higher-point correlators have been extensively investigated in the field-theory limit, corresponding to supergravity or gauge theory in AdS.\footnote{For example, six-point correlators have been studied in~\cite{Alday:2023kfm,Goncalves:2025jcg}, with higher-point supergluon extensions developed in~\cite{Cao:2023cwa,Cao:2024bky}.} Stringy corrections are much less explored, with the first explicit five-point results recently becoming available~\cite{VilasBoas:2025vvw}. We propose a function space of higher-point AdS string building blocks that may be useful for future bootstrap computations at finite $\alpha'$.

These objects should retain the main structural features of flat-space genus-zero string integrals while incorporating the curvature corrections characteristic of AdS. At higher multiplicity, the loci where two punctures collide, \(z_i=z_j\), give rise to logarithmic singularities underlying iterated integrals on the moduli space \(\mathcal M_{0,n}\). This motivates the use of \textit{multivariable} multiple polylogarithms for the AdS insertions, as the natural higher-point extension of the four-point construction. The resulting building blocks define a class of integrals that brings together moduli-space geometry, multivariable polylogarithms and non-commutative monodromy, potentially revealing new relations among special functions and periods.

\subsection{Polylogarithms from the worldsheet}
\label{sec:worldsheet-motivation}

A useful way to motivate the relevant class of functions is to view
curvature corrections to string amplitudes as insertions of additional
gravitons carrying soft momenta.\footnote{We focus on the closed-string
case. For the analogous open-string discussion, see~\cite{Alday:2024ksp}.}
This is illustrated by a toy nonlinear $\sigma$-model whose AdS target
space is expanded around flat space; see, for example,
\cite{Alday:2023jdk,Alday:2023mvu}. Schematically,
\begin{equation}
  G_{\mu\nu}(X)
  =\eta_{\mu\nu}+\frac{1}{R^2}h_{\mu\nu}(X)+\ldots\,,
\end{equation}
where $R$ is the AdS radius. In normal coordinates, the leading
curvature correction is quadratic in $X$ and can be generated by
differentiating twice the plane-wave factor $e^{iq\cdot X}$ of a
graviton vertex, then setting its momentum $q$ to zero.
Evaluating this insertion in the flat-space worldsheet theory
motivates the schematic relation
\begin{equation}
  \delta M^{\rm AdS}_{n}
  \sim
  \left.
  \frac{\partial}{\partial q^\mu}
  \frac{\partial}{\partial q^\nu}
  M^{\rm flat}_{n+1}(p_1,\ldots,p_n,q)
  \right|_{q=0}\,,
  \label{eq:soft-curvature-schematic}
\end{equation}
involving a flat-space amplitude with one additional soft insertion.
This relation is understood formally at the level of the worldsheet
integrand. At higher orders in the curvature expansion, the
$\sigma$-model generates multiple such insertions, alongside
higher-order interaction vertices.

To see how multivariable polylogarithms arise, consider five hard
gravitons with punctures
$(z_1,\ldots,z_5)=(0,x,y,1,\infty)$ and an additional soft graviton
of momentum $q$ at $u$. Two derivatives of the soft Koba--Nielsen
factor $\prod_{i=1}^{4}|u-z_i|^{2\alpha' p_i\cdot q}$
with respect to $q$, evaluated at $q=0$, produce logarithmic
insertions of the form
\begin{equation}
  I_{rs}(x,y;\bar x,\bar y)
  =
  \int_{\mathbb C}d^2u\,
  f(u;x,y)\,\bar f(\bar u;\bar x,\bar y)\,
  \log|u-r|^2\,\log|u-s|^2\,,
  \label{eq:Irs-five-point}
\end{equation}
where $r,s\in\{0,1,x,y\}$ and $f(u;x,y)$ is rational, with poles
at the hard punctures. For fixed $(x,y)$, the soft coordinate $u$
parametrises the punctured sphere
$\mathbb P^1\setminus\{0,x,y,1,\infty\}$, the fibre of the
forgetful map $\mathcal M_{0,6}\to\mathcal M_{0,5}$.
After suitable regularisation, integration over $u$ produces
single-valued multivariable polylogarithms of weight at most three:
two from the logarithmic insertions and one from the integration.

The same reasoning extends to arbitrary multiplicity, with each
soft puncture parametrising the fibre of
$\mathcal M_{0,n+1}\to\mathcal M_{0,n}$
\cite{brown2006multiplezetavaluesperiods}.
It also provides a weight-counting argument. At order $R^{-2k}$,
the normal-coordinate expansion of the $\sigma$-model generates
at most $2k$ powers of $X$, and hence at most $2k$ logarithms.
There are at most $k$ additional punctures to integrate over,
each increasing the polylogarithmic weight by at most one.
Within this class of soft-insertion integrals, the resulting
functions therefore have maximal weight $3k$, independently
of multiplicity, with lower-weight contributions also allowed. This motivates the use of the full generating series of
multivariable (single-valued) multiple polylogarithms to organise
the possible insertions at arbitrary multiplicity.
A complete worldsheet formulation, including curvature corrections
to the action and physical vertex operators, would determine
which insertions in this tower are realised and their coefficients.

Further evidence for the proposed function space comes from
strings in AdS$_3$ with pure NS-NS flux, which admit an exact
WZW worldsheet description~\cite{Maldacena:2000hw}.
Using the screening-integral representation
of~\cite{Alday:2024rjs}, we have verified, through transcendental
weight three, that the corresponding insertions at arbitrary
multiplicity and for any fixed non-negative integer screening
number belong to the space of single-valued multivariable MPLs
on $\mathcal M_{0,n}$. The calculation can be extended recursively to higher weights.
The details will be presented separately.

\subsection{Main results}
Our first main result is an \textit{all-multiplicity} monodromy
relation for the open-string AdS building blocks. In flat space,
monodromy follows from the analytic continuation of the
Koba--Nielsen factor, while the four-point AdS construction
already involves KZ equations and Drinfeld
associators $\Phi$~\cite{Alday:2025cxr}. Beyond four points, however, the extension is far from
automatic: the polylogarithmic insertions become genuinely
multivariable, and their continuation between ordering
chambers requires matching different tangential base-point
normalisations. We derive the corresponding regulator-independent crossing
factors in \eqref{MonodromyMagicFormula} and show how they
combine in the non-commutative order dictated by successive
puncture collisions, as in \eqref{RjAnCont}.

Our result is formulated in terms of the AdS-deformed doubly ordered disc integrals $Z^{(\mathrm{AdS})}(\tau|\rho)$, which generalise the universal building blocks of flat-space open-string amplitudes introduced above. For fixed Parke–Taylor ordering $\rho$, the monodromy relation reads
\begin{equation}
  \sum_{j=1}^{n-1}
  Z^{(\mathrm{AdS})}(\tau_j|\rho)\,
  \mathfrak M_j(s,e;\Phi)
  =0\,,
  \label{eq:main-result-AdS-monodromy}
\end{equation}
where $\tau_j$ label the integration-cycle orderings
specified in Section~\ref{sec:Monodromy Relations}.
The non-commutative monodromy factors are given by
\begin{equation}
\mathfrak M_j=
\overleftarrow{\prod_{k=3}^{j}}
\left[
\Phi\left(e_{2k},\hat\tau_k(e^<_{k-1})\right)
e^{-i\pi (s_{2k}+e_{2k})}
\Phi\left(\hat\tau_{k-1}(e^<_{k-1}),e_{2k}\right)
\right],
\quad j\geq3,
\label{eq:main-monodromy-factors}
\end{equation}
with $\mathfrak M_1=e^{i\pi (s_{12}+e_{12})}$ and
$\mathfrak M_2=1$. Here, $e_m^<$ sums generators with labels smaller than $m$ and $\hat\tau_k$
acts by permuting the generators. The arrow denotes an ordered product. These factors combine the usual Koba--Nielsen phases
with the associator-dependent transport of the
polylogarithmic insertion.

\vspace{0.5em}
Our second main result is an \textit{all-multiplicity} KLT factorisation for the closed-string AdS building blocks. The same crossing data allow the sphere integrals to be expressed in terms of left-moving and appropriately transported right-moving disc building blocks,
\begin{equation}
\begin{aligned}
  J^{(\mathrm{AdS})}(\tau|\rho)
  ={}&\sum_{\alpha,\beta\in S_{n-3}}
  Z^{(\mathrm{AdS})}(\alpha\,1\,n-1\,n|\rho)\,
  S_{(\mathrm{AdS})}[\alpha|\beta]\widetilde Z^{(\mathrm{AdS})}(1\,\beta\,n-1\,n|\tau)\,.
\end{aligned}
\label{eq:main-result-AdS-KLT}
\end{equation}
The kernel $S_{(\mathrm{AdS})}$ is a non-commutative deformation of the flat-space KLT momentum kernel, whose matrix elements factorise into products of sine functions at arbitrary multiplicity~\cite{Bjerrum-Bohr:2010pnr}. Remarkably, this iterative structure persists in AdS, where the crossing factors are non-commutative and their ordering becomes essential. We find
\begin{equation}
  S_{(\mathrm{AdS})}[\alpha|\beta]
  =\left(\frac{i}{2\pi}\right)^{n-3}
  F^{(\alpha)}
  K_{\beta(n-2)}^{(\alpha)}
  \cdots K_{\beta(2)}^{(\alpha)}\,,
  \label{eq:schematic-AdS-kernel-factorisation}
\end{equation}
where the $K_{\beta(k)}^{(\alpha)}$ encode successive contour deformations and $F^{(\alpha)}$ restores the standard tangential base-point normalisation. The complete construction is given in Section~\ref{sec:KLT relations}.

\vspace{0.5em}
At four points, the polylogarithmic insertion depends on a single modulus, and the KLT basis has a single element. The monodromy relation reduces to the result of~\cite{Alday:2025cxr}, as checked at the end of Section~\ref{sec:Monodromy Relations}, while the KLT kernel reduces to $S_{(\mathrm{AdS})}[2|2]=-\pi^{-1}\sin(\pi E_{12})$, reproducing~\cite{Alday:2025bjp}; see \eqref{klt4pt}. Moreover, in the limit $e_{ij}\to0$, the associators trivialise and both relations recover their flat-space counterparts at arbitrary multiplicity. A worked five-point example illustrating the monodromy
and KLT relations is presented in
Appendix~\ref{app:five-point-example}.

We stress that these relations hold at the level of the proposed AdS building blocks; determining their coefficients in physical string amplitudes is a separate dynamical problem.

\begin{table}[htbp]
\centering
\small
\renewcommand{\arraystretch}{1.25}
\begin{tabular}{@{}l@{\hspace{0.7em}}p{0.76\linewidth}@{}}
\toprule
Symbol & Meaning \\
\midrule

$\tau,\rho$
& Disc: integration cycle ($\tau$) and Parke--Taylor ordering ($\rho$).
\newline Sphere: two Parke--Taylor orderings. \\
\addlinespace[3pt]

$Z,\ Z^{(\mathrm{AdS})}$
& Flat-space and MPL-dressed AdS disc integrals. \\
\addlinespace[3pt]

$\widetilde Z^{(\mathrm{AdS})}$
& Transported right-moving disc integral in the AdS KLT relation. \\
\addlinespace[3pt]

$J,\ J^{(\mathrm{AdS})}$
& Flat-space and svMPL-dressed AdS sphere integrals. \\
\addlinespace[3pt]

$G_\tau,\ \mathrm{sv}(G_{\mathbb I})$
& Chamber-normalised MPL series and canonical single-valued series. \\
\addlinespace[3pt]

$e_{ij},\ E_{ij}=s_{ij}+e_{ij}$
& Non-commuting KZ generators and shifted Mandelstam variables. \\
\addlinespace[3pt]

$b_\tau,\ \Phi$
& Tangential base points and Drinfeld associator. \\
\addlinespace[3pt]

$S_{(\mathrm{AdS})}$
& KLT kernel. \\
\bottomrule
\end{tabular}
\caption{Main notation used throughout the paper.}
\label{tab:main-notation}
\end{table}

\paragraph{Outline.} Section~\ref{sec:Multiple polylogarithms} introduces the polylogarithmic functions and conventions used throughout. Section~\ref{sec:Building blocks} reviews the flat-space disc and sphere integrals and proposes their AdS counterparts, along with their properties. Sections~\ref{sec:Monodromy Relations} and~\ref{sec:KLT relations} establish the AdS monodromy and KLT relations, respectively. We conclude with a discussion of open questions. 

\section{Multiple polylogarithms}\label{sec:Multiple polylogarithms}

Multiple polylogarithms (MPLs) can be described as
iterated integrals on punctured Riemann spheres.
The associated non-commutative generating series
satisfy Knizhnik--Zamolodchikov (KZ) equations,
which make their analytic continuation and
monodromy properties particularly transparent.

\paragraph{One-variable generating series.}
Let \(A=\{a_0,a_1,\ldots,a_m\}\) be an alphabet of distinct marked points, and \(e_{a_0},e_{a_1},\ldots,e_{a_m}\) the non-commuting formal variables attached to these labels. We denote by
\begin{equation}
  \mathbb{L}^{A}_{a_0^+}(z)
  :=
  \mathbb{L}_{a_0^+}
  \left[
  \begin{array}{cccc}
    e_{a_0} & e_{a_1} & \ldots & e_{a_m} \\
    a_0 & a_1 & \ldots & a_m
  \end{array}
  ;z
  \right]
\end{equation}
the regularised solution of 
\begin{equation}
  d\mathbb{L}^{A}_{a_0^+}(z)
  =
  \Omega_A(z)\,\mathbb{L}^{A}_{a_0^+}(z) \,,
  \qquad
  \Omega_A(z)
  =
  \sum_{\alpha=0}^{m}
  e_{a_\alpha}\,d\log(z-a_\alpha)\,.
  \label{eq:L-differential-equation}
\end{equation}
Thus, all non-commuting factors act by left multiplication. The subscript \(a_0^+\) denotes a tangential base point at \(a_0\), approached from the
positive real direction. The solution is normalised by the local asymptotic behaviour\footnote{In the following, we will slightly abuse terminology and sometimes refer to such regularised asymptotic normalisations at singular points as boundary values.}
\begin{equation}
  \mathbb{L}_{a_0^+}^A(z)
  \sim
  (z-a_0)^{e_{a_0}}
  \qquad
  \text{as } z\to a_0^+\,,
  \label{eq:tangential-basepoint-asymptotic}
\end{equation}
where $(z-a_0)^{e_{a_0}} :=\exp\!\big(e_{a_0}\log(z-a_0)\big)$.

Away from the singularities \(a_\alpha\), the distinguished solution can be transported along a path \(\gamma\) from a regular reference point \(z_0\) to \(z\), avoiding the marked points in \(A\). The transport is given by the path-ordered exponential
\begin{equation}
  \mathbb{L}_{a_0^+}^A(z)
  =
  \mathcal{P}\exp\!\left(\int_{\gamma}\Omega_A\right)\,
 \mathbb{L}_{a_0^+}^A(z_0)\,.
  \label{eq:path-ordered-solution}
\end{equation}
In our left-action convention, contributions from later points along the path appear to the left of earlier ones. Expanding the path-ordered exponential therefore gives regularised iterated integrals multiplying words in the non-commuting generators \(e_{a_\alpha}\). The generating series takes the form
\begin{equation}
  \mathbb{L}^{A}_{a_0^+}(z)
  =
  \sum_{r=0}^{\infty}\hspace{2pt}
  \sum_{\alpha_1,\ldots,\alpha_r=0}^{m}
  e_{a_{\alpha_1}}\ldots e_{a_{\alpha_{r}}}\,
  L_A(a_{\alpha_1},\ldots,a_{\alpha_r};z)\,.
  \label{eq:L-word-expansion}
\end{equation}
The coefficients \(L_A(a_{\alpha_1},\ldots,a_{\alpha_r};z)\) are the corresponding regularised iterated integrals, i.e. hyperlogarithms (multiple polylogarithms) with letters in \(A\), namely singularities at the marked points in \(A\). The length \(r\) of the word is the transcendental weight. The empty word gives \(L_{A}(\emptyset,z)=1\), while repeated letters at the tangential base point reduce to powers of logarithms, \(L_A(a_0^\ell;z)=\log^\ell(z-a_0)/\ell!\).

We will also need the comparison between regularised solutions at different tangential base points. In the two-letter case, the Drinfeld associator \(\Phi(e_0,e_1)\) is defined by the asymptotic expansion
\begin{equation}
  \mathbb{L}_{0^+}^{\{0,1\}}(z) \sim (1-z)^{e_1}\,\Phi(e_0,e_1)
  \qquad
  \text{as } z\to1^- \,.
  \label{eq:associator-def}
\end{equation}
Equivalently, $\Phi(e_0,e_1)$ is the connection
matrix relating the solutions normalised at
$0^+$ and $1^-$. Its expansion in $e_0,e_1$
has multiple zeta values as coefficients,
and in our conventions
$\Phi(e_0,e_1)^{-1}=\Phi(e_1,e_0)$. We use the real Drinfeld associator.

\paragraph{Multivariable generating series and braid algebra.}

We now turn to the multivariable functions relevant for higher-point
string integrals. These functions are iterated integrals on
\(\mathcal M_{0,n}\), the moduli space of \(n\) ordered marked points
on the Riemann sphere~\cite{brown2006multiplezetavaluesperiods}.

The term ``multivariable'' refers to their dependence on the \(n-3\)
moduli of \(\mathcal M_{0,n}\), rather than on a single complex variable. In the gauge $(z_1,z_{n-1},z_n)=(0,1,\infty)$, these moduli are represented by
\((z_2,\ldots,z_{n-2})\). The finite singular divisors in this affine
chart, corresponding to boundary divisors of
\(\overline{\mathcal M}_{0,n}\), are
\begin{equation}
  z_i=0,
  \qquad
  z_i=1,
  \qquad
  z_i=z_j \,.
  \label{eq:branch-loci}
\end{equation}
Starting at five points, the divisors $z_i=z_j$
can involve two unfixed punctures, introducing
moduli-dependent letters and genuinely
multivariable iterated integrals. Globally on
$\mathcal M_{0,n}$, the divisors $z_i=\infty$
must also be included. Before gauge fixing, we start from the multivariable KZ connection on the configuration space,
\begin{equation}
  \Omega_n
  =
  \sum_{1\leq i<j\leq n}
  e_{ij}\,d\log(z_i-z_j) \,.
  \label{eq:mvKZ-connection}
\end{equation}
The non-commuting variables obey $e_{ij}=e_{ji}$ and satisfy the infinitesimal braid relations 
\begin{equation}
\begin{aligned}
  [e_{ij},e_{kl}]
  &=
  0\,,
  \qquad
  i,j,k,l \text{ all distinct}\,,
  \\
  [e_{ij},e_{ik}+e_{jk}]
  &=
  0\,,
  \qquad
  i,j,k \text{ distinct}\,.
\end{aligned}
\label{eq:braid-algebra}
\end{equation}
These relations are precisely what is needed for the flatness of \(\Omega_n\). They ensure that analytic continuations between chambers are well-defined up to homotopy and can be computed by ordered products of local monodromy factors and associators. We further impose the analogues of the usual Mandelstam constraints\footnote{See equation \eqref{mandelstam}.}
\begin{equation}
  e_{ii}=0\,,
  \qquad
  \sum_{j=1}^{n} e_{ij}=0,
  \qquad
  i=1,\ldots,n \,.
  \label{eq:e-mandelstam-constraints}
\end{equation}
Thus, generators involving the puncture at infinity may be eliminated. The \(i=n\) constraint then gives the residual finite relation \(\sum_{1\le i<j\le n-1}e_{ij}=0\). We will use the shorthands 
\begin{equation}
  e_m^<
  :=
  \sum_{j=1}^{m-1} e_{jm},
  \qquad
  e_m^>
  :=
  \sum_{j=m+1}^{n} e_{mj} \,.
  \label{eq:e-less-greater-def}
\end{equation}
The braid algebra is permutation covariant. We denote by \(\hat\tau\) the induced action on generators, $\hat\tau(e_{ij}):=e_{\tau(i)\tau(j)}$. In the canonical chamber,\footnote{In Section \ref{sec:Monodromy Relations}, this chamber is denoted by \(\mathcal R_2\), because the regions are numbered according to the position of the second puncture.}
\[
\mathcal R_{\mathbb I}
=
\{\,0<z_2<z_3<\cdots<z_{n-2}<1\,\} \,,
\]
we choose the tangential boundary point \(b_{\mathbb I}\) specified by
the nested limit 
\[
0<z_2\ll z_3\ll\cdots\ll z_{n-2}\ll1 \,.
\]
The multivariable generating series normalised at this tangential base
point is
\begin{equation}\label{eq:G new def}
G_{\mathbb I}^{[b_{\mathbb I}]}
(z_2,\ldots,z_{n-2};\{e_{ij}\})
=
\mathbb L_1\mathbb L_2\cdots\mathbb L_{n-3} \,.
\end{equation}
For notational simplicity, we will generally suppress the superscript
\([b_{\mathbb I}]\). 

For \(k=1,\ldots,n-3\), the factor \(\mathbb{L}_k\) is the regularised solution, in the variable \(z_{k+1}\), of
\begin{align}
  d_{z_{k+1}}\mathbb{L}_k
  =
  \bigg[
    e^<_{k+1}\,d\log z_{k+1}
    +
    \sum_{j=k+2}^{n-2}
    e_{j,k+1}\,d\log(z_{k+1}-z_j)
    +
    e_{n-1,k+1}\,d\log(z_{k+1}-1)
  \bigg]\mathbb{L}_k \,,
  \label{eq:Lk-diff-eq}
\end{align}
with asymptotic normalisation
\begin{equation}
  \mathbb{L}_k
  \sim
  z_{k+1}^{\,e^<_{k+1}}
  \qquad
  \text{as } z_{k+1}\to0^+,
  \label{eq:Lk-asymptotic}
\end{equation}
while the variables \(z_{k+2},\ldots,z_{n-2}\) are kept fixed. Equivalently, $\mathbb L_k$ is the one-variable
generating series with alphabet
$A_k=\{0,z_{k+2},\ldots,z_{n-2},1\}$,
and corresponding residues
$e^<_{k+1},e_{k+2,k+1},\ldots,e_{n-1,k+1}$.

The ordered product in \eqref{eq:G new def} gives the fibration-basis
representation of multivariable MPLs. In the
canonical chamber, the larger variables appear
as letters for the smaller ones, reducing analytic
continuation to a sequence of one-variable problems. The choice of chamber is part of the definition of \(G_\tau\). For orderings $\tau$ preserving the fixed punctures at $0$, $1$ and $\infty$, we define the chamber
\begin{equation}
  \mathcal R_\tau
  =
  \{\,0<z_{\tau(2)}<\cdots<
  z_{\tau(n-2)}<1\,\},
  \label{eq:tau-region}
\end{equation}
where the label \(n\) denotes the puncture at infinity and is not included in the real inequalities. The canonical chamber corresponds to \(\mathbb I=(1,2,\ldots,n-1,n)\). The multivariable generating series in the chamber \(\mathcal R_\tau\) is denoted by $G_{\tau}(z_2,\ldots,z_{n-2};\{e_{ij}\})$. We define 
\begin{equation}
  G_{\tau}(z_2,\ldots,z_{n-2};\{e_{ij}\})
  :=
  G_{\mathbb I}
  \big(
    z_{\tau(2)},\ldots,z_{\tau(n-2)};
    \hat{\tau}\{e_{ij}\}
  \big)\,.
  \label{eq:G-tau-def}
\end{equation}
Orderings in which a puncture crosses $0$ or $1$
are treated separately in
Section~\ref{sec:Monodromy Relations}. 

The generating series \(G_\tau\) is normalised at the tangential
boundary point 
\begin{equation}
b_\tau:=\tau\!\cdot b_{\mathbb I} \,,
\label{eq:tangential-basepoints}
\end{equation}
specified by the
nested limit 
\[
0<z_{\tau(2)}\ll z_{\tau(3)}
\ll\cdots\ll z_{\tau(n-2)}\ll1.
\]
As for $G_{\mathbb I}$, we suppress the base-point
superscript $[b_\tau]$ unless otherwise specified.

\paragraph{Single-valued generating series.}
The MPLs introduced above are generally multi-valued. For the proposed AdS closed-string building blocks, we need their single-valued counterparts. The single-valued map \(\operatorname{sv}\) acts coefficient-wise on the non-commutative word expansion, producing real-analytic combinations of holomorphic and anti-holomorphic MPLs whose monodromies cancel~\cite{Brown:2004ugm}. We call these functions single-valued multiple polylogarithms (svMPLs). For example, at weight one, up to the choice of base-point normalisation, \(\operatorname{sv}(L_A(a_0;z))=\log|z-a_0|^2\). 

For the multivariable generating series, we similarly define
\begin{equation}
  \mathcal G_{\tau}
  (z_2,\ldots,z_{n-2};
  \bar z_2,\ldots,\bar z_{n-2};
  \{e_{ij}\})
  :=
  \operatorname{sv}\!\left(G_{\tau}\right)
  (z_2,\ldots,z_{n-2};\{e_{ij}\})\,.
  \label{eq:G-tau-sv-def}
\end{equation}
Its coefficients in a non-commutative word basis are single-valued multivariable MPLs on \(\mathcal M_{0,n}\), with letters associated with the boundary divisors in \eqref{eq:branch-loci}.

To describe their holomorphic--anti-holomorphic
factorisation, let $t$ denote the anti-involution of
the completed non-commutative word algebra,
defined by $(XY)^t=Y^tX^t$ and $e_{ij}^t=e_{ij}$.
We use the shorthand
$G_\tau^t(\bar z_i):=(\overline{G_\tau(z_i)})^t$,
where complex conjugation acts only on the scalar
coefficient functions, while $t$ reverses the order
of the generators in each word. The factorised form of the single-valued generating series depends on
the tangential base point used to normalise \(G_\tau\). We write
\begin{equation}
  \mathcal G_\tau(z_i,\bar z_i)
  =
  G_\tau(z_i)\,
  \mathcal{U}_\tau\,
  G_\tau^t(\bar z_i)\,
  \mathcal{U}_\tau^{-1} \,,
  \label{eq:svG-factorisation}
\end{equation}
where \(\mathcal{U}_\tau\) is independent of the moduli and encodes the
dependence on the tangential
base point \(b_\tau\). Following~\cite{Frost:2023stm}, we denote by
$\operatorname{sv}(\mathbb M)$ the single-valued
generating series of multiple zeta values in the
extended braid algebra. For the canonical
normalisation, $\mathcal U_{\mathbb I}
=\operatorname{sv}(\mathbb M)$; other tangential
base points are related by constant transport.

We conclude by describing the differential equations satisfied by
\(\mathcal G_\tau\). 
Let $\Omega_\tau$ denote the gauge-fixed KZ connection
in the chamber $\mathcal R_\tau$, so that
$dG_\tau=\Omega_\tau G_\tau$.
Complex conjugation followed by the anti-involution $t$
reverses the order of the non-commuting factors, giving
$\bar d G_\tau^t=G_\tau^t\overline{\Omega}_\tau$,
where $\overline{\Omega}_\tau$ is the complex-conjugate
connection. Since $\mathcal U_\tau$ is independent of
the moduli, the factorisation
\eqref{eq:svG-factorisation} then implies
\begin{equation}
  d\,\mathcal G_\tau
  =\Omega_\tau\,\mathcal G_\tau,
  \qquad
  \bar d\,\mathcal G_\tau
  =\mathcal G_\tau
  \left(
    \mathcal U_\tau\,
    \overline{\Omega}_\tau\,
    \mathcal U_\tau^{-1}
  \right).
  \label{eq:sv-holo-antiholo-KZ}
\end{equation}
\section{Building blocks}\label{sec:Building blocks}

We now use the polylogarithmic generating series introduced
in the previous section to construct the AdS building blocks.
We first recall the corresponding flat-space integrals and
their properties, before examining how these are modified
by the polylogarithmic insertions.

\subsection{Flat space}
Tree-level string amplitudes in flat space depend on the
external momenta $p_i^\mu$ and polarisations. We introduce
the dimensionless Mandelstam invariants
$s_{ij}:=2\alpha' p_i\cdot p_j=s_{ji}$.
For massless and momentum-conserving scattering, they satisfy
\begin{equation}\label{mandelstam}
s_{ii}=0,\qquad \sum_{j=1,\dots,n}s_{ij}=0\,,
\end{equation}
leaving $n(n-3)/2$ independent invariants.

As discussed in the Introduction, flat-space open-
and closed-string amplitudes can be expressed in terms
of doubly ordered disc and sphere integrals
~\cite{Broedel:2013tta,Stieberger:2014hba}.
For the disc integrals, $\tau$ specifies the ordering
of the integration cycle and $\rho$ the Parke--Taylor
ordering; for the sphere integrals, both permutations
specify Parke--Taylor factors:
\begin{align}
&\begin{aligned}\label{eq:Open-string BB in flat space}
    Z(\tau|\rho)
    :=
    \int_{\gamma(\tau)}
    \frac{dz_1 dz_2\ldots dz_n}{\mathrm{vol}\,\mathrm{SL}(2,\mathbb{R})}
    \,
    (-1)^{n-3}\,
   \mathrm{KN}_\tau(z)\,
    \mathrm{PT}_{\rho}(z)\,,
\end{aligned}\\[2ex]
&\begin{aligned}\label{eq:Closed-string BB in flat space}
    J(\tau|\rho) := \int_{\mathbb{C}^n} \hspace{2pt} \frac{d^2z_1 d^2 z_2 \ldots d^2 z_n}{\pi^{n-3} \,\text{vol }\text{SL}(2,\mathbb{C})} \, \prod_{1\leq i<j\leq n} |z_{ij}|^{2s_{ij}}\, \text{PT}_\rho(z) \text{PT}_\tau(\bar{z}) \,,
\end{aligned}
\end{align}
Here $z$ collectively denotes the puncture positions. The Koba--Nielsen (KN) and Parke--Taylor (PT) factors are\footnote{For the purposes of Section \ref{sec:Monodromy Relations}, we define the Koba--Nielsen factor as a holomorphic function. This definition agrees with the standard convention involving absolute values.}
\begin{align}
\begin{split}
    \mathrm{KN}_{\tau}(z):=
    \prod_{1 \leq i < j \leq n}(z_{\tau(j)}-z_{\tau(i)})^{s_{\tau(i)\tau(j)}}\,, \qquad 
    \mathrm{PT}_{\rho}(z)
:=
\frac{1}{\prod_{i=1}^{n}z_{\rho(i)\rho(i+1)}}\, ,
    \end{split}
\end{align}
where $z_{ij}:=z_i-z_j$, $\tau,\rho\in S_n$, with $\rho(n+1)=\rho(1)$. The disc integration cycle is
\[
\gamma(\tau)
=
-\infty < z_{\tau(1)} < z_{\tau(2)} < \ldots
< z_{\tau(n-1)} < z_{\tau(n)} < \infty \,.
\]
The sphere integrals instead run over complex puncture positions. In both cases, fixing three punctures to $(z_i,z_j,z_k)=(0,1,\infty)$ removes the $\mathrm{SL}(2,\mathbb R)$ or $\mathrm{SL}(2,\mathbb C)$ redundancy and leaves $n-3$ integration variables. The corresponding Jacobian is $|z_{ij}z_{ik}z_{jk}|$ on the disc and its square on the sphere. For later comparison with their AdS counterparts, we recall
the main relations satisfied by these integrals~\cite{Broedel:2013tta,Schlotterer:2018zce}.

\paragraph{Open strings}

\begin{itemize}[leftmargin=*]
\item \textbf{Cyclicity and reflection}:
The geometry of the integration cycle and the cyclic
structure of the Parke--Taylor factor imply cyclicity
and reflection symmetry in both orderings. For example,
\begin{align}
Z(1\,2\ldots n|\rho)
&=Z(2\,3\ldots n\,1|\rho)\,, \qquad
Z(1\,2\ldots n|\rho)=(-1)^n Z(n\,n-1\ldots 2\,1|\rho)\,.
\label{eq:Open-string reflection}
\end{align}
    
\item \textbf{Integration-by-parts relations}: The vanishing of total derivatives in the punctures, combined with partial-fraction identities, relates different orderings $\rho$ at fixed $\tau$,
\begin{align}\label{eq:Flat-space IBP}
        \sum_{j=2}^{n-1} (s_{12}+s_{13}+\ldots+s_{1j})Z(\tau|2\,3 \ldots j \,1 \,j+1\ldots n-1 \,n)=0 \hspace{20pt}\forall \tau \in S_n\,.
    \end{align}
\end{itemize}
The geometry of the integration domain also provides
a hypergeometric interpretation of the disc integrals.
They are Aomoto--Gelfand hypergeometric functions
associated with the hyperplane arrangement
\eqref{eq:branch-loci}. At multiplicity $n\geq4$, their type $(p+1,q+1)$ is given by
\[
p+1=n-2\,,
\qquad \qquad
q+1
=
2(n-3)+\binom{n-3}{2}\,.
\]
\paragraph{Closed strings.}
Since both orderings label Parke--Taylor factors,
the sphere integrals are symmetric,
$J(\tau|\rho)=J(\rho|\tau)$, and satisfy
cyclicity and integration-by-parts relations
in either ordering. The latter take the same
form as \eqref{eq:Flat-space IBP}, with $Z$
replaced by $J$.

\subsection{AdS}
In gauge-fixed form, the AdS building blocks are constructed from the flat-space integrals \eqref{eq:Open-string BB in flat space}, \eqref{eq:Closed-string BB in flat space} by inserting $(n-3)$-variable MPLs (open strings) and svMPLs (closed strings).\footnote{By promoting the variables in the generating series of $(n-3)$-variable MPLs/svMPLs to cross-ratios, one obtains AdS building blocks before fixing the gauge. In this work, we use their gauge-fixed form.} Their generating series are
\begin{align}
&\begin{aligned}\label{eq:Open-string BB in AdS}
    Z&^{(\text{AdS})}(\tau|\rho;\{e_{ij}\}):= \\[1ex]  & \int_{\gamma(\tau)}\hspace{5pt} \prod_{i=2}^{n-2} dz_{i}\hspace{4pt}(-1)^{n-3} \, \text{KN}_\tau(z) \, \text{PT}_\rho(z)\, G_{\tau}(z_2,\ldots,z_{n-2};\{e_{ij}\}) 
\end{aligned}\\[4ex]
&\begin{aligned}\label{eq:Closed-string BB in AdS}
    J&^{(\text{AdS})}(\tau|\rho;\{e_{ij}\}):=\\[1.5ex] 
    & \frac{1}{\pi^{n-3}} \,\int_{\mathbb{C}^{n-3}} \,\prod_{i=2}^{n-2} \, d^2z_i\, \prod_{1\leq i<j\leq n-1} \hspace{-5pt}|z_{ij}|^{2s_{ij}} \,\text{PT}_\rho(z) \, \text{PT}_\tau(\bar{z})\, \text{sv}(G_{\mathbb{I}})(z_2,\ldots,z_{n-2};\{e_{ij}\}) \,. 
\end{aligned}
\end{align}
The generating series $G_\tau$ and ${\rm sv}(G_{\mathbb I})$
are normalised at the tangential boundary points $b_\tau$
and $b_{\mathbb I}$, respectively, defined in
\eqref{eq:tangential-basepoints}.
For open strings, the chamber of $G_\tau$ must match
the integration cycle $\gamma(\tau)$, and we use the gauge
$(x_{\tau(1)},x_{\tau(n-1)},x_{\tau(n)})=(0,1,\infty)$.
Changes of gauge or base point are implemented through
changes of variables and products of Drinfeld associators.
For closed strings, we choose the canonical single-valued
series ${\rm sv}(G_{\mathbb I})$, although any suitably
regular svMPL insertion is allowed. Individual building blocks are obtained by extracting
the coefficients of words in the braid generators $e_{ij}$,
whose dependence we will generally suppress in the notation.

The polylogarithmic insertions affect the flat-space
relations in different ways. Properties inherited
directly from the Parke--Taylor factors remain unchanged,
while relations involving changes of integration cycle
or derivatives of the integrand can acquire
non-commutative corrections. We discuss these properties
below, leaving monodromy and KLT to Sections
\ref{sec:Monodromy Relations} and
\ref{sec:KLT relations}, respectively.

\paragraph{Open strings}
\begin{itemize}[leftmargin=*]
\item \textbf{Cyclic relations}: The cyclic invariance in $\rho$ is unaffected by the MPL insertion, as it follows directly from the Parke--Taylor factor. In contrast, a cyclic permutation of the integration-cycle ordering $\tau$ requires a Möbius transformation to relate the corresponding gauge-fixed chambers. Comparing the MPL generating series at their tangential base points then yields a product of Drinfeld associators. The change of variables 
\begin{align}
    w_i= 1 - \frac{z_2}{z_i}\,, \hspace{10pt}i=1,2,\ldots,n\,
\end{align}
maps the canonical cycle $(1\,2\ldots n)$ and gauge $(z_1,z_{n-1},z_n)=(0,1,\infty)$ to $(2\,3\ldots n\,1)$ with gauge $(w_2,w_n,w_1)=(0,1,\infty)$. Thus, we obtain
\begin{align}
\begin{split}
&\begin{aligned}
  Z^{(\text{AdS})}&(1\,2\,3 \ldots n|\rho)
  =Z^{(\text{AdS})}(2\,3\ldots n\, 1|\rho)\overleftarrow{\prod_{k=3}^{n-1}}\Phi\left(e_{k}^>,e^<_{k}-e_{1k}\right) \hspace{15pt} \forall \rho \in S_n
\end{aligned}\\[2ex]
&\begin{aligned}
Z^{(\text{AdS})}(\tau|1\,2\,3 \ldots n)=Z^{(\text{AdS})}(\tau|2\,3\ldots n\,1) \hspace{112pt} \forall \tau \in S_n \,,
    \end{aligned}
    \end{split}
\end{align}
where $e_k^<$ and $e_k^>$ are defined in
\eqref{eq:e-less-greater-def}. The arrows indicate
the order of multiplication:
\begin{align}\label{eq:Def product with arrow}
    \overrightarrow{\prod_{k=i}^{j}}\, A_k := A_{i} \,A_{i+1}\ldots A_{j-1}\,A_{j}\hspace{40pt}     \overleftarrow{\prod_{k=i}^{j}}\, A_k := A_{j}\,A_{j-1}\ldots A_{i+1}\,A_{i} \,.
\end{align}
\item \textbf{Reflection identity}: Unlike cyclic permutations of the integration cycle,
reflection does not introduce additional Drinfeld
associators. The Möbius map $w_i=1-z_i$ sends the
canonical cycle to $(n\,n-1\ldots 2\,1)$, with gauge $(w_n,w_{n-1},w_1)=(-\infty,0,1)$. We find
\begin{equation}
Z^{(\mathrm{AdS})}(1\,2\ldots n|\rho)
=(-1)^n
Z^{(\mathrm{AdS})}(n\ldots2\,1|\rho),
\end{equation}
and similarly for reflection in $\rho$.

\item \textbf{Integration-by-parts relations}:
The Koba--Nielsen factor and the MPL generating series
obey KZ-type equations with residues $s_{ij}$ and
$e_{ij}$, respectively. Therefore,
\begin{equation}
\partial_{z_i}\bigl(\mathrm{KN}_\tau G_\tau\bigr)
=
\left(\sum_{k\neq i}\frac{E_{ik}}{z_{ik}}\right)
\mathrm{KN}_\tau G_\tau,
\qquad E_{ij}:=s_{ij}+e_{ij}.
\end{equation}
The flat-space IBP relation
\eqref{eq:Flat-space IBP} thus extends to AdS
with the replacement $s_{ij}\to E_{ij}$:
\begin{align}
            \sum_{j=2}^{n-1} (E_{12}+E_{13}+\ldots+E_{1j})Z^{({\rm AdS})}(\tau|2\,3\ldots j \,1 \,j+1\ldots n-1 \,n)=0 \hspace{15pt} \forall \tau \in S_n\,,
\end{align}
At four points, for example,\vspace{-1pt}
\begin{align}\label{eq:AdS IBP at four points} E_{12}\,Z^{(\text{AdS})}(\tau|2\,1\,3\,4)+(E_{12}+E_{13})Z^{(\text{AdS})}(\tau|2\,3\,1\,4)=0\,.
\end{align}
Applying the transposition $\sigma=(1\,2)$ and choosing $\tau=(1\,2\,3\,4)$, one recovers the relations under shifts in the Mandelstam $s := s_{12}$ first shown in~\cite{Alday:2025bjp}.

\item \textbf{Derivative relations}:
Differentiating the Koba--Nielsen factor with respect to
$s_{ij}$ produces a logarithmic insertion, which can
equivalently be generated by acting on the MPL series.
Let $D_{ij}$ be the formal derivation on the
non-commutative word algebra, defined by
$D_{ij}e_{kl}=\delta_{ij,kl}$ and the Leibniz rule,
before imposing relations among the generators.
The shuffle identities of iterated integrals then imply\footnote{At fixed word \(w\), this means that \(\partial_{s_{ij}}\) produces the sum of all insertions obtained by adding one letter \(e_{ij}\) anywhere in \(w\). See \cite{Alday:2025bjp} for the four-point case.}
\begin{align}\label{eq:Open-string derivative relations}
\partial_{s_{ij}}
Z^{(\mathrm{AdS})}(\tau|\rho;\{e_{kl}\})
=
D_{ij}
Z^{(\mathrm{AdS})}(\tau|\rho;\{e_{kl}\})\,,
\qquad \forall\,\tau,\rho\in S_n\,.
\end{align}
\end{itemize}
As at four points~\cite{Alday:2025bjp}, the AdS open-string integrals are Aomoto--Gelfand hypergeometric functions, with type enlarged by the MPL insertion. At $n\geq4$, their type $(p+1,q+1)$ is 
\begin{align}
    p+1= n+r-2, \qquad  \qquad q+1= 2(n-3) + \binom{n-3}{2}+r+d \,,
\end{align}
where $r$ and $d$ are the respective sums of the weights and depths of the one-variable MPLs constituting the multivariable MPL in the open-string integral.

\paragraph{Closed strings}
\begin{itemize}[leftmargin=*]
\item \textbf{Cyclicity and symmetry}:
Since both orderings enter through the Parke--Taylor
factors, cyclic invariance is unaffected by the AdS
insertion. Exchanging $\tau$ and $\rho$, however,
also involves complex conjugation of the
single-valued insertion:\footnote{Here the bar acts coefficient-wise as defined in Section~\ref{sec:Multiple polylogarithms}, exchanging \(z_i\) and \(\bar z_i\) while leaving the non-commutative words unchanged.}
\begin{equation}
J^{(\mathrm{AdS})}\!\left(\tau\middle|\rho;\{e_{ij}\}\right)
=
\left.
J^{(\mathrm{AdS})}\!\left(\rho\middle|\tau;\{e_{ij}\}\right)
\right|_{\mathcal{G}_{\mathbb{I}}\,\to\,\overline{\mathcal{G}_{\mathbb{I}}}} \, .
\end{equation}

\item \textbf{Integration-by-parts relations}: The closed-string IBP relations follow from the
holomorphic and anti-holomorphic KZ equations
\eqref{eq:sv-holo-antiholo-KZ}, with non-commuting
coefficients acting from the left and right,
respectively:
\begin{align}
\begin{split}
&\begin{aligned}
        \sum_{j=2}^{n-1} (E_{12}+E_{13}+\ldots +E_{1j})J^{(\text{AdS})}(\tau|2\,3\ldots j \,1 \,j+1 \ldots n-1 \,n)=0 \hspace{15pt} \forall \tau \in S_n 
\end{aligned}\\[1.5ex]
&\begin{aligned}
        \sum_{j=2}^{n-1} J^{(\text{AdS})}(2\,3\ldots j \,1 \,j+1\ldots n-1 \,n|\rho) (\tilde{E}_{12}+\tilde{E}_{13}+\ldots+\tilde{E}_{1j})=0\hspace{15pt} \forall \rho \in S_n\,.
        \end{aligned}
        \end{split}
\end{align}
Here, $\tilde{E}_{ij}:=\operatorname{sv}(\mathbb M)\,
E_{ij}\operatorname{sv}(\mathbb M)^{-1}$.

\item \textbf{Derivative relations}: The closed-string derivative relations are the single-valued analogues of the open-string ones. Let \(D^{\mathrm{sv}}_{ij}\) denote the single-valued lift of the derivation \(D_{ij}\), acting on the single-valued generating series \(\mathcal G_\tau(z,\bar z;e)=\mathrm{sv}(G_\tau)\). The closed-string AdS building blocks obey
\begin{align}\label{eq:Closed-string derivative relations}
\partial_{s_{ij}}
J^{(\mathrm{AdS})}(\tau|\rho;\{e_{kl}\})
=
D^{\mathrm{sv}}_{ij}
J^{(\mathrm{AdS})}(\tau|\rho;\{e_{kl}\}) \,.
\end{align}
\end{itemize}
We now turn to relations involving analytic continuation
between integration chambers, where the non-commutative
transport of the polylogarithmic insertions plays a central role.

\section{Monodromy relations}\label{sec:Monodromy Relations}

Flat-space monodromy relations constrain tree-level
colour-ordered amplitudes and their $Z$-theory building
blocks~\cite{Plahte:1970wy,Bjerrum-Bohr:2009ulz,Mafra:2016mcc}.
A non-commutative AdS generalisation was recently derived
at four points~\cite{Alday:2025cxr}.
In this section, we extend the monodromy relations to non-commutative \(Z\)-theory building blocks at arbitrary multiplicity.

The proof involves analytically continuing the integrand of the building blocks and integrating along a closed contour. We first work in the canonical chamber. From Cauchy's theorem,

\begin{multline}
    \int_{\gamma({\mathbb{I}_{n-1})}} \prod_{i=3}^{n-2} dx_i \int_{-\infty}^{+\infty}dx_2\,
    \mathrm{KN}_{\mathbb{I}}(x_2+i\varepsilon,x_3,\ldots,x_{n-2})
    \,\mathrm{PT}_{\rho}(x) G_{\mathbb{I}}(x_2+i\varepsilon,x_{3},\ldots,x_{n-2})=0\,,
    \label{Cauchy}
\end{multline}
where $\gamma(\mathbb I_{n-1})$ is the cycle defined by
$0<x_3<\cdots<x_{n-2}<1$, and the insertion $G_{\mathbb I}$
is the generating series of multivariable MPLs given by
\eqref{eq:G new def}. After gauge fixing,
$\mathrm{KN}_{\mathbb I}\mathrm{PT}_\rho$ behaves as
$x_2^{-s_{2n}-1}$ or $x_2^{-s_{2n}-2}$, depending on $\rho$,
while the MPL insertion grows at most as powers of
$\log|x_2|$ at fixed weight. We can therefore drop the arc
at infinity in a suitable convergence domain and close
the contour in the upper half-plane, where there are no
singularities. The resulting identities extend by analytic
continuation in the Mandelstam variables. We denote the
real integration variables by $x_i$.

Since $G_{\mathbb I}$ is normalised in the canonical
chamber $0<x_2<x_3<\ldots<x_{n-2}<1$, the key step
is to determine its analytic continuation to
$x_2+i\varepsilon$ in the other ordering chambers. We decompose the integration domain in \eqref{Cauchy} according to the position of \(x_2\) relative to the ordered points $(0,x_3,\ldots,x_{n-2},1)$:
\begin{align}
\sum_{k=1}^{n-1}\int_{\mathcal{R}_k} \left (\prod_{i=2}^{n-2} dx_i\right)
\mathrm{KN}_{\mathbb{I}}(x_2+i\varepsilon,& x_3,\ldots,x_{n-2})
\mathrm{PT}_{\rho}(x)
G_{\mathbb{I}}(x_2+i\varepsilon,x_{3},\ldots,x_{n-2})
=0 \,,
\label{Cauchy2}
\end{align}
where the orderings $\tau_k$ are defined by
\begin{align}
    &\tau_{1} = (2\,1\,3\ldots n-2\,n-1\,n)\,, \qquad \tau_2\equiv\mathbb{I}_n\,, \nonumber
    \\[0.4em]
    &\tau_{j} = (1\,3\ldots j\,2\,j+1\ldots n-2\,n-1\,n)\,,
    \quad 2<j\leq n-1\nonumber
\end{align}
and $\mathcal{R}_j$ is the region associated with the
ordering $\tau_j$. For $j\neq2$, we need to relate
the analytic continuation of $G_{\mathbb{I}}$ to the
generating series $G_{\tau_j}$ normalised in the
corresponding chamber. To this end, we use the KZ equation:
\begin{align}
    G_{\mathbb{I}}(z_2,\ldots,z_{n-2})=\mathcal{P}\mathrm{exp}\left(\int_{x_j\leadsto z_j}\Omega\right)G_{\mathbb{I}}(x_2,\ldots,x_{n-2})\,,\label{pathOrderedContG}
\end{align} 
where $x_j\rightsquigarrow z_j$ denotes a path in the
complexified configuration space, starting from the
real chamber $\mathcal R_2$, ending in $\mathcal R_j$,
and avoiding the singular divisors of $\Omega$.
We choose this path so that the transport factorises
into elementary monodromy phases and Drinfeld
associators. To express the result in terms of the
chamber-normalised series $G_{\tau_j}$, we must
compare the regularised limits across successive
collision divisors. The auxiliary regulators then
cancel, and the crossing factors combine in the
non-commutative order prescribed by the path.
We do this case by case.
\subsection*{Analytic continuation to $\mathcal{R}_1$}
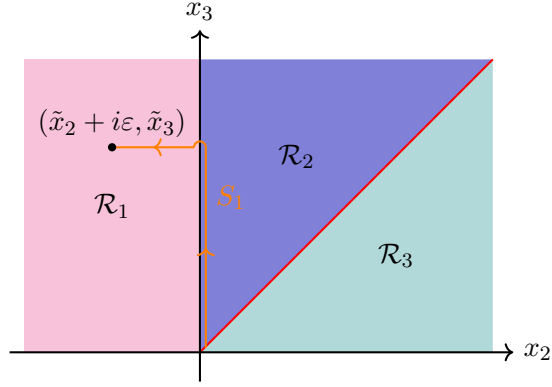
\begin{figure}[h!]
            \centering
            \resizebox{0.5\textwidth}{!}{
            \begin{tikzpicture}[scale=4]
                \def\xmax{1.0}
                \def\ymax{1}


                \fill[magenta!60, opacity=0.5]
                    (-0.6, 0) rectangle (0, \ymax);

                \fill[blue!70!black, opacity=0.5]
                    (0,0) -- (0,1) -- (1,1) -- cycle;

                \fill[teal!60, opacity=0.5]
                    (0,0) -- (1,0) -- (1,1) -- cycle;

                \draw[red,thick] (0,0) -- (1, 1);

                \draw[->, thick] (-0.65, 0) -- (\xmax+.07, 0) node[right] {$x_2$};
                \draw[->, thick] (0, -0.1) -- (0, \ymax+.1) node[above] {$x_3$};

                \draw[orange, thick, postaction={decorate, decoration={markings, mark=at position 0.5 with {\arrowreversed{>}}}}] (-0.3, 0.7) -- (-0.02, 0.7);
                \draw[orange, thick] (0.02, 0.7) arc (0:180:0.02);
                \draw[orange, thick, postaction={decorate, decoration={markings, mark=at position 0.5 with {\arrowreversed{>}}}}] (0.02, 0.7) -- (0.02, 0.015)
                    node[pos=0.25, right, orange] {$S_1$};

                \fill[black] (-0.3, 0.7) circle (0.4pt);
                \node[above] at (-0.3, 0.7) {$(\tilde{x}_2+i\varepsilon,\tilde{x}_3)$};

                \node at (-0.3, 0.5) {$\mathcal{R}_{1}$};
                \node at (0.33, 0.67) {$\mathcal{R}_{2}$};
                \node at (0.67, 0.33) {$\mathcal{R}_{3}$};
            \end{tikzpicture}
            }
            \caption{Contour used to define the analytic continuation of $G_{\mathbb{I}}$ just above $\mathcal{R}_1$. The small arc around the branch point \(x_2=0\) goes up in the $\mathrm{Im}(x_2)$ direction to avoid the branch line.}
            \label{ContourR1}
        \end{figure}

This region is associated with the ordering $\tau_1=(2\,1\,3\ldots n)$. Since there is only one branch line separating $\mathcal{R}_1$ and $\mathcal{R}_2$, this is the region whose choice of path will be the simplest. We start from $(\epsilon_2,\epsilon_3,x_4,\ldots,x_{n-2})\in \mathcal{R}_2$ and move the first two coordinates along the path $S_1$ in Figure \ref{ContourR1} to $(\tilde x_2+i\varepsilon,\tilde x_3)$.
Computing the path-ordered exponential along $S_1$ gives
\begin{align}
&G_{\mathbb{I}}(x_2+i\varepsilon,x_3,\ldots,x_{n-2}) =
\notag\\
{}&
\tilde G_{\tau_1}(x_2,\ldots,x_{n-2})
\tilde G^{-1}_{\tau_1}(-\epsilon_2,x_3,\ldots,x_{n-2})\
e^{i\pi e_{12}}
G_{\mathbb{I}}(\epsilon_2,x_3,\ldots,x_{n-2}) \,,
\end{align}
where $\tilde G_{\tau_1}$ is a MPL generating series which is well-defined in $\mathcal{R}_1$, given by
\begin{align}
\tilde{G}_{\tau_1}\equiv\mathbb{L}_{0^-}\!\left[\begin{smallmatrix} e_{12} & e_{23} & \ldots&e_{2,n-3}&e_{2,n-2}&e_{2,n-1}\\ 0 & x_3 &\ldots &x_{n-3}&x_{n-2}&1\end{smallmatrix}; x_{2}\right]\mathbb{L}_{2}\ldots \mathbb{L}_{n-3}\,.\label{tau1MPL}
\end{align}
From \eqref{tau1MPL} we have that 
\begin{align}
\tilde G^{-1}_{\tau_1}(-\epsilon_2,x_3,\ldots,x_{n-2})
e^{i\pi e_{12}}
G_{\mathbb{I}}(\epsilon_2,x_3,\ldots,x_{n-2})
&=
\mathbb L_{n-3}^{-1}\!\ldots\!\mathbb L_2^{-1}
\epsilon_2^{-e_{12}}e^{i\pi e_{12}}\epsilon_2^{e_{12}}
\mathbb L_2\!\ldots\!\mathbb L_{n-3}
\nonumber\\
&=e^{i\pi e_{12}} \,,
\end{align}
where the braid relations imply that \([\mathbb L_k,e_{12}]=0\) for \(k>1\). Therefore,
\begin{align}
    G_{\mathbb{I}}(x_2+i\varepsilon,x_3,\ldots,x_{n-2})=\tilde G_{\tau_1}e^{i\pi e_{12}}\qquad \mathrm{for}\;\{x_i\}\in\mathcal{R}_1.\label{R1AnCont}
\end{align}
Finally, to relate the integral on \(\mathcal
{R}_1\) to \(Z^{\mathrm{(AdS)}}(\tau_1|\rho)\), we perform the change of variables
\begin{equation}
    x_2=\frac{x_1^\prime}{x^\prime_{1}-1},\qquad x_j=\frac{x_j^\prime-x_1^\prime}{1-x^\prime_{1}},\qquad j=3,\ldots,n-2,
\end{equation}
and use that
\begin{equation*}
    \tilde{G}_{\tau_{1}}\left(\frac{x_1^\prime}{x^\prime_{1}-1},\frac{x_3^\prime-x_1^\prime}{1-x^\prime_{1}},\ldots ,\frac{x_{n-2}^\prime-x_1^\prime}{1-x^\prime_{1}};\{e_{ij}\}\right)=G_{\mathbb{I}}(x_1^\prime,x_3^\prime,\ldots,x_{n-2}^\prime,\{\hat\tau_{1}\cdot e_{ij}\})
\end{equation*}
to infer that the contribution of \(\mathcal{R}_1\) to (\ref{Cauchy2}) is
\begin{equation}
    Z^{\mathrm{(AdS)}}(\tau_1|\rho)e^{i\pi E_{12}}\, ,\label{monTerm-1}
\end{equation}
where the phase in the Mandelstam variable comes from the usual analytic continuations of the KN factor.
\subsection*{Analytic continuation to $\mathcal{R}_j$, with $2<j\leq n-2$}

Once again, we need a convenient choice of path that makes the path-ordered exponentials computable. The first step is to define the boundary points $\mathcal{D}^\pm_j$. Given the regions $\mathcal{R}_{j-1}$ and $\mathcal{R}_{j}$, we call $\mathcal{D}_j$ the $x_{2}=x_j$ diagonal interface connecting them. Then $\mathcal{D}_j^{-(+)}$ is a point in $\mathcal{R}_{j-1(j)}$ close to the interface. More precisely, $\mathcal{D}^-_j$ is defined by taking $(x_2,\ldots, x_{n-2})\approx(\epsilon_2,\ldots,\epsilon_{n-2})$ with 
\begin{equation}
  \epsilon_3 \ll \epsilon_4 \ll \ldots \ll
  \epsilon_{j-1} \ll \epsilon_2 \approx \epsilon_j
  \ll \epsilon_{j+1} \ll \ldots \ll \epsilon_{n-2} \ll 1\,,
  \label{eq:Dj-minus-eps-ordering}
\end{equation}
where by $\epsilon_2\approx\epsilon_j$ we mean that 
\begin{equation}
    \epsilon_2/\epsilon_j=1-\gamma\qquad \mathrm{with}\;0<\gamma\ll1\,.\label{RatioCond}
\end{equation}
The point $\mathcal{D}_j^+$ is obtained by reflecting $\mathcal{D}_j^-$ across the interface $\mathcal{D}_j$ in the $(x_2,x_j)$ plane.
  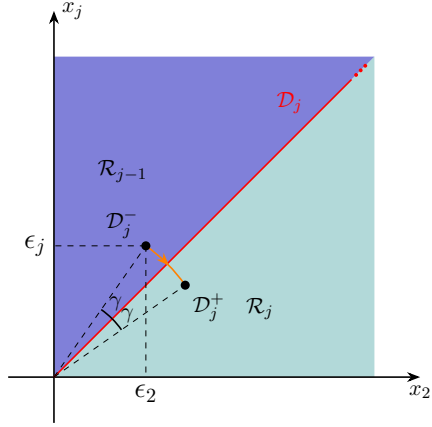
\begin{figure}[h!]
            \centering
            \resizebox{0.39\textwidth}{!}{
            \begin{tikzpicture}[scale=0.8,
                    >=Stealth,
                    dot/.style={circle, fill=black, inner sep=0pt, minimum size=4.5pt},
                    midarrow/.style={decoration={markings, mark=at position #1 with {\arrow{>}}}, postaction={decorate}}
                ]
                \fill[blue!70!black, opacity=0.5]
                    (0,0) -- (0, 7) -- (7, 7) -- cycle;
                \node at (1.5, 4.5) {$\mathcal{R}_{j-1}$};
                \fill[teal!60, opacity=0.5]
                    (0,0) -- (7, 0) -- (7, 7) -- cycle;
                \node at (4.5, 1.5) {$\mathcal{R}_{j}$};
                \draw[->, thick] (-1, 0) -- (8, 0) node[below] {$x_2$};
                \draw[->, thick] (0, -1) -- (0, 8) node[right] {$x_j$};
                \draw[red, thick] (0,0) -- (6.5, 6.5);
                \fill[red] (6.6, 6.6) circle (1.2pt);
                \fill[red] (6.7, 6.7) circle (1.2pt);
                \fill[red] (6.8, 6.8) circle (1.2pt);
                \node[red, above left] at (5.6, 5.6) {$\mathcal{D}_j$};
                \def\redangle{45}
                \def\deltaangle{10}
                \def\pOneAngle{\redangle + \deltaangle}
                \def\pTwoAngle{\redangle - \deltaangle}
                \def\radius{3.5}
                \coordinate (P1) at (\pOneAngle:\radius);  
                \coordinate (P2) at (\pTwoAngle:\radius);  
                \draw[dashed] (0,0) -- (P1);
                \draw[dashed] (0,0) -- (P2);
                \draw[dashed] (P1) -- (0,0 |- P1);
                \draw[dashed] (P1) -- (P1|-0,0);
                \draw[thick] (\redangle:1.8) arc (\redangle:\pOneAngle:1.8);
                \node at ({\redangle + \deltaangle/2}:2.1) {$\gamma$};
                \draw[thick] (\pTwoAngle:1.8) arc (\pTwoAngle:\redangle:1.8);
                \node at ({\redangle - \deltaangle/2}:2.1) {$\gamma$};
                \draw[orange, thick, midarrow=0.55] (\pOneAngle:\radius) arc (\pOneAngle:\pTwoAngle:\radius);
                \node[dot] at (P1) {};
                \node[dot] at (P2) {};
                \node[above left] at (P1) {$\mathcal{D}^-_j$};
                \node[below right] at (P2) {$\mathcal{D}^+_j$};
                \node[left, font=\Large] at (0,0|-P1) {$\epsilon_j$};
                \node[below, font=\Large] at (P1|-0,0) {$\epsilon_2$};
            \end{tikzpicture}
            }
           \caption{The diagonal $\mathcal{D}_j$ separates $\mathcal{R}_{j-1}$ from $\mathcal{R}_j$ in the $(x_2,x_j)$ plane. Analytic continuation across this branch line is performed by a small arc in the $\mathrm{Im}(x_2)$ direction, from $\mathcal{D}_j^-$ to $\mathcal{D}_j^+$ (orange path).}
            \label{GeomConfigDj} 
        \end{figure}

There are two auxiliary results we need to prove before obtaining the full analytic continuation formula: the asymptotic behaviours of $G_{\tau_{j-1}}$ in $\mathcal{D}_{j}^-$ and of $G_{\tau_{j}}$ in $\mathcal{D}_{j}^+$.
Since $G_{\tau_{j-1}}$ is as defined in \eqref{eq:G-tau-def}, we can write it in $\mathcal{D}^{-}_j$ as
\begin{align}
    G_{\tau_{j-1}}^{\mathcal{D}^-_j}&\sim\left(\prod_{k=3}^{j-1}\epsilon_k^{\hat\tau_{j-1}(e^{<}_{k-1})}\right)    \mathbb{L}_{0^+}\!\left[\begin{smallmatrix} \hat\tau_{j-1}(e^<_{j-1}) & e_{2j}\\ 0 & \epsilon_j \end{smallmatrix}; \epsilon_2\right]\left(\prod_{l=j}^{n-2}\epsilon_l^{e_l^{<}}\right)
    \nonumber\\&
    =\left(\prod_{k=3}^{j-1}\epsilon_k^{\hat\tau_{j-1}(e^{<}_{k-1})}\right)    \mathbb{L}_{0^+}\!\left[\begin{smallmatrix} \hat\tau_{j-1}(e^<_{j-1}) & e_{2j}\\ 0 & 1\end{smallmatrix}; 1-\gamma\right]\epsilon_j^{\hat\tau_{j-1}(e_{j-1}^{<})+e_j^<}\left(\prod_{l=j+1}^{n-2}\epsilon_l^{e_l^{<}}\right)
    \nonumber\\&
    \sim \left(\prod_{k=3}^{j-1}\epsilon_k^{\hat\tau_{j-1}(e^{<}_{k-1})}\right)\gamma^{e_{2j}}\Phi(\hat\tau_{j-1}(e^<_{j-1}),e_{2j})\epsilon_j^{\hat\tau_{j-1}(e_{j-1}^{<})+e_j^<}\left(\prod_{l=j+1}^{n-2}\epsilon_l^{e_l^{<}}\right).
\end{align}
The first and third relations are asymptotic equalities in the hierarchy of regulators and in the limit \(\gamma\to0^+\), respectively. The equality in the second line follows from the exact rescaling identity
\[
\mathbb L_{0^+}\!\left[
\begin{smallmatrix}
e_{0x} & e_{xy}\\
0 & y
\end{smallmatrix}
;x
\right]
=
\mathbb L_{0^+}\!\left[
\begin{smallmatrix}
e_{0x} & e_{xy}\\
0 & 1
\end{smallmatrix}
;x/y
\right]y^{e_{0x}}\,,
\]
together with \eqref{RatioCond}. The combination of the \(\epsilon_j\)-powers uses commutation relations, which we now justify. It remains to show that we can commute all regulators to the left of the Drinfeld associator. The first identity to use is
\begin{align}
    [e_l^<,e_{ij}]=0 \qquad i,j<l,\quad i\ne j\,,
\end{align}
which follows trivially from the braid relations. Then,
\begin{align}
    G_{\tau_{j-1}}^{\mathcal{D}^-_j}&\sim\left(\prod_{k=3}^{j-1}\epsilon_k^{\hat\tau_{j-1}(e^{<}_{k-1})}\right)\left(\prod_{l=j+1}^{n-2}\epsilon_l^{e_l^{<}}\right)\gamma^{e_{2j}}\Phi(\hat\tau_{j-1}(e^<_{j-1}),e_{2j})\epsilon_j^{\hat\tau_{j-1}(e_{j-1}^{<})+e_j^<} \,.
\end{align}
To commute the remaining regulator, we use permutation covariance of the braid algebra:
\begin{align}
    [a,b]=c \iff [\hat\sigma( a),\hat\sigma(b)]=\hat\sigma( c)\,.
\end{align}
Here $\sigma$ is any element of $S_n$. We obtain
\begin{align}
    \hat\tau_{j-1}^{-1}\!\left([\hat\tau_{j-1}(e_{j-1}^{<})+e_j^<,e_{2j}]\right)&=[e^<_{j-1}+e_j^{<},e_{j-1,j}]\nonumber\\
    &=\left[\sum_{k=1}^{j-2}(e_{k,j-1}+e_{kj})+e_{j-1,j},e_{j-1,j}\right] \\
    & =0\,.
\end{align}
Thus, from covariance
\begin{equation}
     [\hat\tau_{j-1}(e_{j-1}^{<})+e_j^<,e_{2j}]=0\,.
\end{equation}
Let us use the same argument for the last commutator:
\begin{align}
    \hat\tau_{j-1}^{-1}\!\left([\hat\tau_{j-1}(e_{j-1}^{<})+e_j^<,\hat\tau_{j-1}(e_{j-1}^<)]\right)&=[ e_{j-1}^{<}+e_j^<, e_{j-1}^<]
    =0\,.
\end{align}
Again, covariance implies
\begin{align}
    [\hat\tau_{j-1}(e_{j-1}^{<})+e_j^<,\hat\tau_{j-1}(e_{j-1}^<)]=0 \,.
\end{align}
This allows us to write
\begin{align}
    G_{\tau_{j-1}}^{\mathcal{D}^-_j}&\sim\left(\prod_{k=3}^{j-1}\epsilon_k^{\hat\tau_{j-1}(e^{<}_{k-1})}\right)\left(\prod_{l=j+1}^{n-2}\epsilon_l^{e_l^{<}}\right)\gamma^{e_{2j}}\epsilon_j^{\hat\tau_{j-1}(e_{j-1}^{<})+e_j^<}\Phi(\hat\tau_{j-1}(e^<_{j-1}),e_{2j})\, .
    \label{DjMinus}
\end{align}
Since $\mathcal{D}_j^+$ is defined as the reflection of $\mathcal{D}_j^-$ around the $x_2=x_j$ line, we have 
\begin{align}
    G_{\tau_j}^{\mathcal{D}^+_j}=G_{\tau_{j-1}}^{\mathcal{D}^-_j}\vert_{e_{2k}\leftrightarrow e_{jk}}\,.
\end{align}
All regularisation-dependent terms in \eqref{DjMinus} are invariant under $e_{2k}\leftrightarrow e_{jk}$. This is trivial for the first three terms, and the exponent in the fourth term can be written as 
\[
\hat\tau_{j-1}(e_{j-1}^{<})+e_j^<
=
e_{2j}
+
\sum_{\substack{k=1\\ k\neq 2}}^{j-1}
\left(e_{2k}+e_{jk}\right) \,,
\]
which manifests the invariance. Therefore,
\begin{align}
     G_{\tau_j}^{\mathcal{D}^+_j}=\left(\prod_{k=3}^{j-1}\epsilon_k^{\hat\tau_{j-1}(e^{<}_{k-1})}\right)\left(\prod_{l=j+1}^{n-2}\epsilon_l^{e_l^{<}}\right)\gamma^{e_{2j}}\epsilon_j^{\hat\tau_{j-1}(e_{j-1}^{<})+e_j^<}\Phi(\hat\tau_j(e^<_{j-1}),e_{2j})\,.
\end{align}
This gives us the key identity:
\begin{equation}
    \left(G_{\tau_j}^{\mathcal{D}^+_j}\right)^{-1}e^{-i\pi e_{2j}}G_{\tau_{j-1}}^{\mathcal{D}^-_j}=\Phi(e_{2j},\hat\tau_j(e^<_{j-1}))e^{-i\pi e_{2j}}\Phi(\hat\tau_{j-1}(e^<_{j-1}),e_{2j})\,.\label{MonodromyMagicFormula}
\end{equation}
All regularisation factors cancel.

We can now specify the path used to analytically continue $G_{\mathbb{I}}$ from region $2$ to region $j$: we start from $\mathcal{D}_3^-$, cross to $\mathcal{D}_3^+$ along a small arc in the $\mathrm{Im}(x_2)$ direction, and go to the next diagonal. Then, we keep going across diagonals until we reach $\mathcal{R}_j$. Schematically, our path is
\begin{align}
    \mathcal{D}^-_3\to \mathcal{D}^+_3\to\mathcal{D}^-_4\to\mathcal{D}^+_4\to\ldots\to \mathcal{D}^-_{j}\to \mathcal{D}^+_j\,.
\end{align}
The path-ordered exponential along this path gives
\begin{equation}
\begin{aligned}
G_{\mathbb{I}}(x_2+i\varepsilon,x_3,\ldots,x_{n-2})
={}&
G_{\tau_j}(x_2,\ldots,x_{n-2})
\bigl(G^{\mathcal D^+_{j}}_{\tau_j}\bigr)^{-1}
e^{-i\pi e_{2j}}
G^{\mathcal D^-_{j}}_{\tau_{j-1}}
\\[0.4em]
&\times
\bigl(G^{\mathcal D^+_{j-1}}_{\tau_{j-1}}\bigr)^{-1}
e^{-i\pi e_{2,j-1}}
G^{\mathcal D^-_{j-1}}_{\tau_{j-2}}
\ldots
\bigl(G^{\mathcal D^+_{3}}_{\tau_3}\bigr)^{-1}
e^{-i\pi e_{23}}
G^{\mathcal D^-_{3}}_{\mathbb{I}} \,,
\end{aligned}
\end{equation}
for $\{x_i\}\in \mathcal{R}_j$.
Then, we can use \eqref{MonodromyMagicFormula} to get

\begin{align}
    &G_{\mathbb{I}}(x_2+i\varepsilon,x_3,\ldots,x_{n-2})=
    \nonumber\\[0.6em]& 
    G_{\tau_j}(x_2,\ldots,x_{n-2})\Phi(e_{2j},\hat\tau_j(e^<_{j-1}))e^{-i\pi e_{2j}}\Phi(\hat\tau_{j-1}(e^<_{j-1}),e_{2j})\ldots \Phi(e_{23},e_{13})e^{-i\pi e_{23}}\Phi(e_{12},e_{23})
    \nonumber\\[0.5em]&
    =G_{\tau_j}(x_2,\ldots,x_{n-2})\overleftarrow{\prod_{k=3}^{j}}\left(\Phi(e_{2k},\hat\tau_k(e^<_{k-1}))e^{-i\pi e_{2k}}\Phi(\hat\tau_{k-1}(e^<_{k-1}),e_{2k})\right) , \quad \mathrm{for}\; \{x_i\}\in \mathcal{R}_j, \label{RjAnCont}
\end{align}
where we recall that the arrow indicates that the product is ordered from right to left. Therefore, \(\mathcal{R}_j\) contributes to (\ref{Cauchy2}) with
\begin{equation}
   Z^{\mathrm{(AdS)}}(\tau_j|\rho)\overleftarrow{\prod_{k=3}^{j}}\left(\Phi(e_{2k},\hat\tau_k(e^<_{k-1}))e^{-i\pi E_{2k}}\Phi(\hat\tau_{k-1}(e^<_{k-1}),e_{2k})\right)\, .\label{monTerm_j}
\end{equation}
\subsection*{Analytic continuation to $\mathcal{R}_{n-1}$}
The ordering here is $\tau_{n-1}=(1\,3\,4\ldots n-1\,2\,n)$, and the MPL that naturally emerges from the analytic continuation of $G_{\mathbb{I}}$ is
\begin{align}
    \tilde{G}_{\tau_{n-1}}\equiv\hat\tau_{n-1}\!\left(\mathbb{L}_1\ldots\mathbb{L}_{n-4}\right)\mathbb{L}_{1^+}\!\left[\begin{smallmatrix} \sum^{n-2}_{j=1}e_{2j} & e_{2,n-1}\\ 0 &  1\end{smallmatrix}; x_2\right].
\end{align}
We still need to understand how to reach $\mathcal{R}_{n-1}$. The main additional complication has already been addressed. We use our diagonal-by-diagonal construction to reach $\mathcal{R}_{n-2}$ and then let $x_2$ approach $1$ while keeping all other variables very close to zero. As in the other cases, the regularisation factors cancel, and we get
\begin{align}
    &G_{\mathbb{I}}(x_2+i\varepsilon,x_3,\ldots,x_{n-2})=\nonumber\\
    &\tilde G_{\tau_{n-1}}e^{-i\pi e_{2,n-1}}\Phi\!\left(\sum_{j=1}^{n-2}e_{2j},e_{2,n-1}\right)\overleftarrow{\prod_{k=3}^{n-2}}\left[\Phi\!\left(e_{2k},\hat\tau_k(e^<_{k-1})\right)e^{-i\pi e_{2k}}\Phi\!\left(\hat\tau_{k-1}(e^<_{k-1}),e_{2k}\right)\right],\label{Rn-1AnCont}
\end{align}
when $(x_2,\ldots,x_{n-2})\in\mathcal{R}_{n-1}$, and we omit the dependence on $(x_i)$ on the right-hand side. To relate the integral on $\mathcal{R}_{n-1}$ to $Z^{(\mathrm{AdS})}(\tau_{n-1}|\rho)$, we need to map the region to 
\begin{equation}
    0<x^\prime_{3}<\ldots<x^\prime_{n-1}<1.
\end{equation}
To do so, we perform the change of variables
\begin{equation}
    x_2=\frac{1}{x^\prime_{n-1}},\qquad x_j=\frac{x_j^\prime}{x^\prime_{n-1}},\qquad j=3,\ldots,n-2.
\end{equation}
Then, the identity
\begin{equation*}
    \tilde{G}_{\tau_{n-1}}\left(\frac{1}{x^\prime_{n-1}},\frac{x^\prime_{3}}{x_{n-1}^\prime},\ldots ,\frac{x^{\prime}_{n-2}}{x^\prime_{n-1}};\{e_{ij}\}\right)=G_{\mathbb{I}}(x_3^\prime,x_4^\prime,\ldots,x_{n-1}^\prime,\{\hat\tau_{n-1}\cdot e_{ij}\})\Phi(e_{2,n-1},\hat\tau_{n-1}(e_{n-2}^<))
\end{equation*}
together with (\ref{Rn-1AnCont}) implies that this region contributes to (\ref{Cauchy2}) with the term 
\begin{equation}
    Z^{(\mathrm{AdS})}(\tau_{n-1}|\rho)\overleftarrow{\prod_{k=3}^{n-1}}\left(\Phi(e_{2k},\hat\tau_k(e^<_{k-1}))e^{-i\pi E_{2k}}\Phi(\hat\tau_{k-1}(e^<_{k-1}),e_{2k})\right)\, . \label{monTerm_n-1}
\end{equation}
\subsection*{The relations}
Plugging \eqref{monTerm-1}, \eqref{monTerm_j}, and \eqref{monTerm_n-1} in \eqref{Cauchy2}, we find
\begin{align}
    &Z^{(\mathrm{AdS})}(\tau_1|\rho)\,e^{i\pi E_{12}}
    +Z^{(\mathrm{AdS})}(\mathbb{I}_n|\rho)\nonumber\\
    &+\sum_{j=3}^{n-1}Z^{(\mathrm{AdS})}(\tau_j|\rho)
    \overleftarrow{\prod_{k=3}^{j}}
    \left[\Phi\!\left(e_{2k},\hat\tau_k(e^<_{k-1})\right)
    e^{-i\pi E_{2k}}
    \Phi\!\left(\hat\tau_{k-1}(e^<_{k-1}),e_{2k}\right)\right]=0\,.
    \label{MonodromyRelationsI}
\end{align}
These are the monodromy relations for $n$-point non-commutative AdS $Z$-theory integrals. 

A few comments are in order. First, we have performed contour deformations in the positive $\mathrm{Im}(x_2)$ region. If we had chosen $\mathrm{Im}(x_2)=-\varepsilon$,  we would have obtained the complex-conjugate relations. Moreover, given $\sigma \in S_{n}$, we can use permutation covariance to write monodromy relations for any $Z^{(\mathrm{AdS})}(\sigma|\rho)$. To this end, let us define the symbol $C^{(L)}_{k\to j}$ as the factor produced when a positive point $k$ is crossed past point $j$ from left to right, where the set $L$ collects the non-negative points lying to the left of $k$. It is given by 
\begin{align}
    C^{(L)}_{k\to j}=\Phi\!\left(e_{jk},\,\sum_{l\in L}e_{jl}\right)e^{-i\pi E_{jk}}\,\Phi\!\left(\sum_{l\in L}e_{kl},\,e_{jk}\right).
\end{align}
For completeness and later convenience, we also introduce $C^{(R)}_{j\leftarrow k}$, the factor for crossing a negative point $k$ past $j$ from right to left, and the set $R$ collects the non-positive points to the right of $k$. In complete analogy, this reads 
\begin{align}\label{Rcross}
    C^{(R)}_{j \leftarrow k}=\Phi\left(e_{jk},\sum_{r\in R}e_{jr}\right)e^{i\pi E_{jk}}\Phi\left(\sum_{r\in R}e_{kr},e_{jk}\right).
\end{align}
We can then write a monodromy relation for any $\sigma$ as 
\begin{align}
    &Z^{(\mathrm{AdS})}(\sigma(2)\,\sigma(1)\ldots\sigma(n) |\rho)\,e^{i\pi E_{\sigma(1)\sigma(2)}}
    +Z^{(\mathrm{AdS})}(\sigma|\rho)\nonumber\\[0.35em]
    &+\sum_{j=3}^{n-1}Z^{(\mathrm{AdS})}(\sigma(1)\,\sigma(3)\ldots\sigma(j)\,\sigma(2)\,\sigma(j+1)\ldots\sigma(n)|\rho)
    \overleftarrow{\prod_{k=3}^{j}}
    C_{\sigma(2)\rightarrow \sigma(k)}^{(\sigma(1)\,\sigma(3)\ldots\sigma(k-1))}=0\,.
    \label{MonodromyRelations}
\end{align}
Finally, it is an important cross-check that in the commutative limit (i.e. $e_{ij}\to 0$), we recover the $n$-point flat-space monodromy relations \cite{Bjerrum-Bohr:2009ulz}. Setting $n=4$ and considering the canonical ordering in the equation above, we get
\begin{align}
    Z^{(\mathrm{AdS})}(2134|\rho)e^{i\pi E_{12}}+Z^{\mathrm{(AdS)}}(1234|\rho)+Z^{\mathrm{(AdS)}}(1324|\rho)\Phi(e_{23},e_{13})e^{-i\pi E_{23}}\Phi(e_{12},e_{23})=0\,,
\end{align}
which, for \(\rho=(1\,2\,3\,4)\), agrees with the relation of \cite{Alday:2025cxr}.


\section{KLT relations}\label{sec:KLT relations}
We now use the crossing factors derived in the previous
section to establish KLT relations for the closed-string
AdS building blocks. The starting point is the
holomorphic--anti-holomorphic factorisation of the
single-valued generating series \eqref{eq:svG-factorisation}.
In the canonical normalisation, we consider the integral
\ie
&\ma{I}=\frac{1}{\pi^{n-3}}\int_{\mathbb{C}^{n-3}} d^2z_2 \ldots d^2z_{n-2}
\prod_{1 \le i<j \le n-1}|z_{ij}|^{2s_{ij}} \,f(z_i)\,g(\bar{z}_i)\,
G_{\mathbb I}(z_i)\,\text{sv}(\mathbb M)\,
G^t_{\mathbb I}(\bar z_i)\,\text{sv}(\mathbb M)^{-1} \,,
\fe
where \(f(z_i)\) and \(g(\bar z_i)\) are single-valued meromorphic insertions with trivial monodromy. Here, they are chosen to be Parke--Taylor factors.\footnote{Besides PT factors, they can also be other functions coming from the OPE of vertex operators. See \cite{Britto:2021prf} for some examples. Thus, we keep the notation $f(z)$ and $g(\bar{z})$ to emphasise this.} To turn this factorisation of the integrand into a
relation between sphere and disc integrals, we follow
the contour deformation of
\cite{Dotsenko:1984nm,Dotsenko:1984ad,Bjerrum-Bohr:2010pnr}.
Writing $z_i=v_i^1+iv_i^2$, we perform the continuation
\begin{align}
v_i^2 \quad \longrightarrow \quad ie^{-2i\epsilon}v_i^2\,,
\end{align}
where $\epsilon>0$ is small. Working to first
order in $\epsilon$ and introducing
$v_i^\pm=v_i^1\pm v_i^2$, we obtain
\ie
&z_i\to v_i^-+2i\epsilon v_i^2
\qquad
\bar z_i\to v_i^+-2i\epsilon v_i^2\,.
\fe
This continuation allows us to treat $z_i$ and $\bar z_i$
as independent variables, and the integral becomes
\ie
\ma{I}=
\int \prod_{i=2}^{n-2}\left(\frac{i\,dv_i^+dv_i^-}{2\pi}\right)
& \prod_{1\leq i<j\leq n-1}
(v_i^+-v_j^+-2i\epsilon (v_i^2-v_j^2))^{s_{ij}}
(v_i^--v_j^-+2i\epsilon(v_i^2-v_j^2))^{s_{ij}}\\
&\times f(v^-_i)\,g(v^+_i)\,
G_{\mathbb I}\!\left(v_i^-+2i\epsilon v_i^2\right)
\,\operatorname{sv}(\mathbb M)\,
G_{\mathbb I}^{t}\!\left(v_i^+-2i\epsilon v_i^2\right)
\,\operatorname{sv}(\mathbb M)^{-1}
\fe
with $v_1^2=v_{n-1}^2=0$. The $\epsilon$ prescription determines the integration contour.

We first consider the \(v_i^+\) integrations. The \(i\epsilon\) prescription determines on which side of the real axis the branch points lie. For \(v_i^+\) outside \((0,1)\), the corresponding \(v_i^-\) contour can be closed without enclosing singularities, and its contribution vanishes. Thus, the remaining domain is \(0<v_i^+<1\), for \(i=2,\ldots,n-2\), and we decompose it into chambers labelled by a permutation \(\beta\),
\[
  0<v_{\beta(2)}^+<v_{\beta(3)}^+<\ldots
  <v_{\beta(n-2)}^+<1\, .
\]
Equivalently, the associated ordering is \((1\,\beta(2)\ldots\beta(n-2)\,n{-}1\,n)\). The factorised representation of the single-valued insertion requires compatible branches and normalisations of its holomorphic and anti-holomorphic factors. Unlike in flat space, changing the chamber therefore also changes the factorisation matrix. The canonical generating series is related to that in the \(\beta\) chamber by $G_{\mathbb I}=G_{1,\beta,n-1,n}\,\mathcal T_{\beta\to\mathbb I}$, where \(\mathcal T_{\beta\to\mathbb I}\) is the constant KZ comparison series between the two chamber normalisations. It is obtained from the MPL part of the crossing factors discussed in Section~\ref{sec:Monodromy Relations}, namely by omitting their Koba--Nielsen phases. For brevity, we define
\begin{equation}
\mathcal U_\beta
:=
\mathcal T_{\beta\to\mathbb I}\,
\operatorname{sv}(\mathbb M)\,
\overline{\mathcal T_{\beta\to\mathbb I}}^{\,t}\,,
\label{eq:U-beta-def}
\end{equation}
which is the factorisation matrix \(\mathcal U_\tau\) of
\eqref{eq:svG-factorisation} for the chamber
\(\tau=(1\,\beta\,n{-}1\,n)\). The canonical single-valued insertion can therefore be written in this chamber as
\begin{equation}
\begin{aligned}
&G_{\mathbb I}\,\operatorname{sv}(\mathbb M)\,
G^t_{\mathbb I}\,\operatorname{sv}(\mathbb M)^{-1} =
G_{1,\beta,n-1,n}\,\mathcal U_\beta\,
G^t_{1,\beta,n-1,n}\,
\operatorname{sv}(\mathbb M)^{-1}.
\end{aligned}
\label{eq:svG-beta-factorisation}
\end{equation}
The final factor \(\operatorname{sv}(\mathbb M)^{-1}\) retains the canonical normalisation, since this expression represents the original canonical single-valued insertion. Thus, different \(v_i^+\)-chambers lead to different factorisation matrices \(\mathcal U_\beta\). These matrices encode the chamber-dependent
transport of the single-valued pairing and enter
the definition of the right-moving building blocks
in the KLT relation.  
\begin{figure}[t]
\centering
\includegraphics[width=12cm]{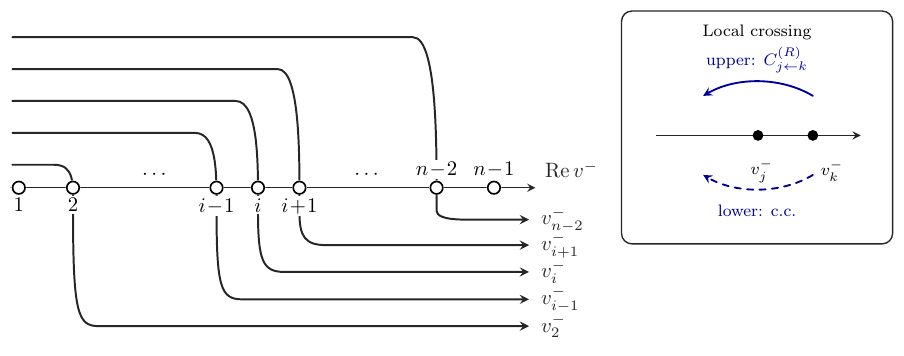}
\caption{Nested integration contours for the variables \(v_i^-\), for the ordering \(0<v_2^+<v_3^+<\cdots<v_{n-2}^+<1\) of the \(v_i^+\) variables. The inset shows the local crossing of \(v_k^-\) past \(v_j^-\), including the extra monodromy factor induced by the polylogarithmic insertion. This picture is inspired by \cite{Bjerrum-Bohr:2010pnr}.}
\label{fig:contourNested}
\end{figure}

We first consider $0<v_2^+<v_3^+<\cdots<v_{n-2}^+<1$. The corresponding contour is shown in Figure~\ref{fig:contourNested}. The \(v_i^-\)-dependent generating series is understood as the analytic continuation of \(G_{\mathbb I}\) from the canonical chamber. To simplify the successive continuations, we temporarily normalise the intermediate generating series at a tangential base point at \(z_1\). For an ordering \((i_1,\ldots,i_m,1,j_1,\ldots,j_k)\), this base point is specified by
\[
|z_{1i_m}|\ll|z_{1i_{m-1}}|\ll\cdots\ll|z_{1i_1}|
\ll|z_{1j_1}|\ll\cdots\ll|z_{1j_k}| \,.
\]
We denote this temporary choice by the superscript \([z_1]\). Once all contours have been deformed, we restore the standard chamber base point \(b_\tau=\tau\!\cdot b_{\mathbb I}\) of \eqref{eq:tangential-basepoints}. The corresponding constant base-point transport \(F^{(\alpha)}\) appears in the KLT kernel below. 

We now focus on the \(v_i^-\) integrations, deforming each \(v_i^-\) contour so that it closes to the left. We begin with the \(v_2^-\) contour. The contribution from the two sides of the cut is
\ie
&\frac{i}{2\pi}\int_{-\infty}^{0} dv_2^-e^{i\pi s_{12}}f(v_i^-)\text{KN}_{(2134\ldots)}
G^{[z_1]}_{(2134\ldots)}(v_i^-)e^{i\pi e_{12}}\\
&-\frac{i}{2\pi}\int_{-\infty}^{0} dv_2^-e^{-i\pi s_{12}}f(v_i^-)\text{KN}_{(2134 \ldots)}
G^{[z_1]}_{(2134\ldots)}(v_i^-)e^{-i\pi e_{12}}\\
&=-\frac{1}{\pi}\int_{-\infty}^{0} dv_2^-f(v_i^-)\text{KN}_{(2134\ldots)}
G^{[z_1]}_{(2134\ldots)}(v_i^-)\text{sin}(\pi E_{12}) \,.
\fe
Here the subscript of \(G\) denotes the chamber ordering. The factor \(e^{i\pi e_{12}}\) accounts for the mismatch between the MPL naturally defined in the ordering \((2\,1\,3\,4\ldots n)\) and the analytic continuation from the ordering \((1\,2\ldots n)\), exactly as in the monodromy relations. For $v_3^-$, there are two contributions: $v_2^-<v_3^-<0$ and $v_3^-<v_2^-<0$. For the region \(v_2^-<v_3^-<0\), the \(v_3^-\) integration gives
\ie
-\frac{1}{\pi}
\int_{v_2^-}^{0} dv_3^-\,
f(v_i^-)\,\mathrm{KN}_{(23145\ldots)}\,
G^{[z_1]}_{(23145\ldots)}(v_i^-)\,
\sin\!\big(\pi E_{13}\big)\Phi(e_{12},e_{13}) \,.
\fe
Here, we display only the factor arising from the \(v_3^-\) integration. Note that the factor $\Phi(e_{12},e_{13})$ comes from the cancellation of the regularisation factors.\footnote{This factor can also be regarded as the result of a base-point change, since the regularisation factors are related to the choice of base point.} Including also the factor from the \(v_2^-\) integration, the full prefactor for this colour ordering, which is part of the KLT matrix element, is
\ie
\frac{1}{\pi^2}
\Phi(e_{13},e_{23})\sin\!\big(\pi E_{13}\big)\Phi(e_{12},e_{13})
\sin\!\big(\pi E_{12}\big) \,.
\fe
Here \(\Phi(e_{13},e_{23})\) restores the standard chamber normalisation, converting \(G^{[z_1]}_{(23145\ldots)}\) into \(G_{(23145\ldots)}\), whose standard base-point label is suppressed according to the convention of Section~\ref{sec:Multiple polylogarithms}. For the $v_3^-<v_2^-<0$ contribution, we get
\ie
&\frac{i}{2\pi}\int_{-\infty}^{v_2^-}dv_3^-f(v_i^-)\text{KN}_{(32145\ldots)}
G^{[z_1]}_{(32145\ldots)}(v_i^-)
\Bigl[\Phi(e_{23},e_{12})e^{i\pi (s_{23}+e_{23})}\Phi(e_{13},e_{23})e^{i\pi (s_{13}+e_{13})}\\
&-\Phi(e_{23},e_{12})e^{-i\pi (s_{23}+e_{23})}\Phi(e_{13},e_{23})e^{-i\pi (s_{13}+e_{13})}\Bigr]\Phi(e_{12},e_{13}) \,.
\fe
Since the Drinfeld associator is real in our conventions, complex conjugation acts only on the explicit phases. The result can then be written as
\ie
&\frac{i}{2\pi}\int_{-\infty}^{v_2^-}dv_3^-f(v_i^-)\text{KN}_{(32145\ldots)}
G^{[z_1]}_{(32145\ldots)}(v_i^-)
\Bigl[\Phi(e_{23},e_{12})e^{i\pi E_{23}}\Phi(e_{13},e_{23})e^{i\pi E_{13}}-\text{c.c.}\Bigr]\Phi(e_{12},e_{13})\,,
\fe
where ``c.c.'' denotes complex conjugation. The order of the non-commuting factors is not reversed.

The contour deformation has a natural iterative structure. Once the ordering of the \(v_i^+\) variables is fixed to be the canonical one, the \(v_i^-\) contours can be deformed one at a time. The deformation of \(v_j^-\) produces an elementary crossing factor \(K_j^{(\alpha)}\), which depends on the position of \(j\) in the left ordering \(\alpha\), relative to the previously deformed labels \(2,\ldots,j-1\). Since the MPL monodromies are in general non-commutative, the order of these elementary factors is important, and the kernel factorises as
\ie
S\left[
\alpha(2)\,\alpha(3)\ldots\alpha(n-2)\,1\,n{-}1\,n
\,\middle|\,
1\,2\ldots n{-}1\,n
\right]
=
\left(\frac{i}{2\pi}\right)^{n-3}
F^{(\alpha)}K_{n-2}^{(\alpha)}
K_{n-3}^{(\alpha)}
\ldots
K_2^{(\alpha)} \,.
\fe
The factor \(F^{(\alpha)}\) restores the standard base-point normalisation after all contour deformations, as explained above. For \(K_2^{(\alpha)}\), we have
\ie
K_2^{(\alpha)}
=
e^{i\pi E_{12}}-e^{-i\pi E_{12}}
=
2i\sin(\pi E_{12}) \,.
\fe
For \(K_3^{(\alpha)}\), the deformation of the \(v_3^-\) contour has two contributions. The first corresponds to the region in which \(3\) remains to the right of \(2\), while the second arises when \(3\) crosses \(2\). Whenever a point \(k\) crosses a previously deformed point \(j\), with both lying on the negative real axis, the crossing contributes the factor given in \eqref{Rcross}. Therefore, when the point \(3\) crosses the point \(2\), we find
\ie
C^{(1)}_{2\leftarrow 3}
=
\Phi(e_{23},e_{12})e^{i\pi E_{23}}\Phi(e_{13},e_{23})\,,
\fe
implying that
\ie
K_3^{(\alpha)}
=
\Big[
\theta_{\alpha}(3,2)
+
C_{2\leftarrow3}^{(1)}\theta_{\alpha}(2,3)
\Big]
e^{i\pi E_{13}}\Phi(e_{12},e_{13})
-\text{c.c.}
\fe
Here, \(\theta_\alpha(i,j)=1\) if \(i\) lies to the right of \(j\) in the left ordering \(\alpha=(\alpha(2)\,\alpha(3)\, \ldots \alpha(n-2))\) of the KLT kernel, and
\(\theta_\alpha(i,j)=0\) otherwise.

For $K_4^{(\alpha)}$, we must distinguish the relative ordering of \(2\) and \(3\) leading to the crossing factors
\ie
\begin{aligned}
C_{3\leftarrow4}^{(1)}
&=
\Phi(e_{34},e_{13})e^{i\pi E_{34}}\Phi(e_{14},e_{34})\,,
\\[6pt]
C_{2\leftarrow4}^{(3,1)}
&=
\Phi(e_{24},e_{12}+e_{23})e^{i\pi E_{24}}
\Phi(e_{14}+e_{34},e_{24})\,,
\\[6pt]
C_{2\leftarrow4}^{(1)}
&=
\Phi(e_{24},e_{12})e^{i\pi E_{24}}\Phi(e_{14},e_{24})\,,
\\[6pt]
C_{3\leftarrow4}^{(2,1)}
&=
\Phi(e_{34},e_{13}+e_{23})e^{i\pi E_{34}}
\Phi(e_{14}+e_{24},e_{34}) \,.
\end{aligned}
\fe
The final expression for $K_4^{(\alpha)}$ is
\ie
K_4^{(\alpha)} =&
\Big[
\theta_{\alpha}(3,2)
\big(
  \theta_{\alpha}(4,3)
  + C_{3\leftarrow4}^{(1)}\theta_{\alpha}(3,4)\theta_{\alpha}(4,2)
  + C_{2\leftarrow4}^{(3,1)}C_{3\leftarrow4}^{(1)}\theta_{\alpha}(2,4)
\big)e^{i\pi E_{14}}\\
&\times\Phi(e_{13},e_{14})\Phi(e_{12}+e_{23},e_{14}+e_{34})\\
&
+
\theta_{\alpha}(2,3)
\big(
  \theta_{\alpha}(4,2)
  + C_{2\leftarrow 4}^{(1)}\theta_{\alpha}(2,4)\theta_{\alpha}(4,3)
  + C_{3\leftarrow4}^{(2,1)}C_{2\leftarrow4}^{(1)}\theta_{\alpha}(3,4)
\big)e^{i\pi E_{14}}\\
&\times\Phi(e_{12},e_{14})\Phi(e_{13}+e_{23},e_{14}+e_{24})\Big]
-\text{c.c.}
\fe
For the general formula, we keep the right ordering canonical and make the dependence on the left ordering \(\alpha\) explicit. The general result follows the same pattern: permutations \(\sigma\) enumerate the possible orderings of the previously deformed contours, while the \(\theta_\alpha\) factors select those compatible with the final left ordering \(\alpha\). Each factor \(C\) accounts for an elementary crossing of the contour of \(k\) past an earlier puncture. Let $\sigma=(\sigma_2\,\ldots\,\sigma_{k-1})$ be a permutation of the labels \((2\,\ldots\,k-1)\), and define
\[
\Theta_\alpha(\sigma)
:=
\prod_{r=2}^{k-2}
\theta_\alpha(\sigma_{r+1},\sigma_r) \,.
\]
With the factor \(\theta_\alpha(k,\sigma_{j-1})\) understood to be \(1\) when \(j=2\), we have
\begin{equation}
\begin{aligned}
K_k^{(\alpha)}
={}&
\Biggl[
\sum_{\sigma}
\Theta_\alpha(\sigma)
\biggl\{
\theta_\alpha(k,\sigma_{k-1})+
\sum_{j=2}^{k-1}
\prod_{\ell=j}^{\overrightarrow{k-1}}
C_{\sigma_\ell\leftarrow k}^{(\sigma_{\ell+1},\ldots,\sigma_{k-1},1)}
\theta_\alpha(\sigma_j,k)\,
\theta_\alpha(k,\sigma_{j-1})
\biggr\}
e^{i\pi E_{1k}}\\
&\times\prod_{j=2}^{\overleftarrow{k-1}}\Phi(e_{1\sigma_j}+\sum_{\ell=j+1}^{k-1}e_{\sigma_j\sigma_\ell},e_{1k}+\sum_{\ell=j+1}^{k-1}e_{\sigma_\ell k})\Biggr]
-\text{c.c.}
\end{aligned}
\label{eq:Kk}
\end{equation}
We now return to a general \(v_i^+\)-chamber labelled by \(\beta\). This does not introduce new crossing factors; it only permutes the order in which the \(v_i^-\) contours are deformed. Thus, the same elementary factors appear in the permuted order, and the general KLT matrix element is
\ie
\begin{aligned}
S_{(\mathrm{AdS})}[\alpha|\beta]
&\equiv
S[\alpha(2)\,\alpha(3)\ldots\alpha(n-2)\,1\,n{-}1\,n
\,|\,
1\,\beta(2)\,\beta(3)\ldots\beta(n-2)\,n{-}1\,n]
\\
&=
\left(\frac{i}{2\pi}\right)^{n-3}
F^{(\alpha)}
K_{\beta(n-2)}^{(\alpha)}
K_{\beta(n-3)}^{(\alpha)}
\cdots
K_{\beta(2)}^{(\alpha)}.
\end{aligned}
\fe
For a general right ordering \(\beta\), the \(v_i^-\) contours are
deformed in the order
\(\beta(2),\ldots,\beta(n-2)\). Accordingly, \(K_{\beta(k)}^{(\alpha)}\) is obtained from \eqref{eq:Kk} by applying \(\beta\) to every puncture label, while leaving all dummy summation and product indices unchanged. All relative positions are evaluated in the fixed left ordering \(\alpha\). From \cite{Drinfeld:1989st,PMIHES_2010__112__143_0} and the definition of Drinfeld associators, the factor $F^{(\alpha)}$ reads
\ie
F^{(\alpha)}=\prod_{j=3}^{\overleftarrow{n-2}}\Phi(e_{1\alpha(j)}+\sum_{\ell=j+1}^{n-2}e_{\alpha(j)\alpha(\ell)},\sum_{\ell=2}^{j-1}e_{\alpha(\ell)\alpha(j)}) \,.
\fe
In the language of AdS $Z$-theory introduced in Section \ref{sec:Building blocks}, the closed-string building blocks factorise as
\begin{align}\label{eq:AdS KLT}
J^{\rm(AdS)}(\tau|\rho)=\sum_{\alpha,\beta\in S_{n-3}}Z^{\rm(AdS)}(\alpha\,1\,n-1\,n|\rho)\,S_{\rm (AdS)}[\alpha|\beta]\,\tilde{Z}^{\rm(AdS)}(1\,\beta\,n-1\,n|\tau)\,.
\end{align}
The first factor is the holomorphic open-string building block defined in \eqref{eq:Open-string BB in AdS}. The corresponding right-moving, or anti-holomorphic, factor is obtained by applying the single-valued conjugation induced by the canonical insertion to the word-valued open-string building block in the chamber \(\beta\), thereby transporting it to the canonical normalisation \(\mathbb I\). At the level of the MPL insertion, this amounts to
\begin{equation}
G_{(1\,\beta\,n{-}1\,n)}(z)
\longmapsto
\mathcal{U}_\beta\,
G^{t}_{(1\,\beta\,n{-}1\,n)}(\bar z)\,
\operatorname{sv}(\mathbb M)^{-1}.
\label{eq:right-moving-G-def}
\end{equation}
We therefore define the transported right-moving block
\begin{equation}
\widetilde Z^{(\mathrm{AdS})}
\bigl(1\,\beta\,n{-}1\,n\mid\tau\bigr)
:=
\mathcal{U}_\beta\,
\bigl[
Z^{(\mathrm{AdS})}
\bigl(1\,\beta\,n{-}1\,n\mid\tau\bigr)
\bigr]^{t}\,
\operatorname{sv}(\mathbb M)^{-1}.
\label{eq:right-moving-Z-def}
\end{equation}
Here, \(t\) reverses the order of the non-commuting words, as in Section~\ref{sec:Multiple polylogarithms}; the anti-holomorphic dependence originates from the conjugate variables and contour prescription.

In the limit $e_{ij}\to0$, the associators and
base-point transport factors become trivial,
and we recover the flat-space KLT kernel
of~\cite{Bjerrum-Bohr:2010pnr}.

\paragraph{Four-point case.}
Let us check this result in the four-point case. Since the KLT basis has a single element, the relation reduces to
\begin{equation}
  J^{(\mathrm{AdS})}(\tau|\rho)
  =
  Z^{(\mathrm{AdS})}(2\,1\,3\,4|\rho)\,
  S_{(\mathrm{AdS})}[2|2]\,
  \widetilde Z^{(\mathrm{AdS})}(1\,2\,3\,4|\tau)\,,
  \label{klt4pt}
\end{equation}
with $ S_{(\mathrm{AdS})}[2|2] = -\pi^{-1}\sin(\pi E_{12})$. In this case, the relevant monodromy relations for the holomorphic block are
\begin{equation}
\begin{aligned}
  &Z^{(\mathrm{AdS})}(2\,1\,3\,4|\rho)\,
  \sin(\pi E_{12})
  =
  Z^{(\mathrm{AdS})}(1\,3\,2\,4|\rho)\,
  \Phi(e_{23},e_{13})\sin(\pi E_{23})\,
  \Phi(e_{12},e_{23})\,,
  \\
  &Z^{(\mathrm{AdS})}(1\,2\,3\,4|\rho)
  =
  -Z^{(\mathrm{AdS})}(2\,1\,3\,4|\rho)\,
  \sin(\pi E_{12})
  \left[
    \cot(\pi E_{12})
    +
    \Phi(e_{23},e_{12})\,
    \cot(\pi E_{23})\,
    \Phi(e_{12},e_{23})
  \right] .
\end{aligned}
\label{eq:four-point-Z-reduction}
\end{equation}
Substituting these relations into \eqref{klt4pt}, we obtain the canonical-basis form
\begin{equation}
  J^{(\mathrm{AdS})}(\tau|\rho)
  =
  Z^{(\mathrm{AdS})}(1\,2\,3\,4|\rho)\,
  \mathcal K\,
  \widetilde Z^{(\mathrm{AdS})}(1\,2\,3\,4|\tau)\,,
\end{equation}
with
\begin{equation}
  \mathcal K
  = \pi^{-1}
  \left(
    \cot(\pi E_{12})
    +
    \Phi(e_{23},e_{12})\,
    \cot(\pi E_{23})\,
    \Phi(e_{12},e_{23})
  \right)^{-1}.
\end{equation}
This agrees with the result of \cite{Alday:2025bjp}, upon identifying
\(\widetilde Z^{(\mathrm{AdS})}\) with the right-moving block used there.
In our conventions, this block is obtained by the right-moving map
\eqref{eq:right-moving-G-def}, which is the multivariable analogue of the alphabet change in the four-point
construction.
\section{Conclusions} 

In this paper, we proposed open- and closed-string
AdS building blocks at arbitrary multiplicity,
motivated by recent progress on string corrections
to holographic CFT correlators. They are obtained by inserting
multivariable multiple polylogarithms and their
single-valued analogues into the standard disc and sphere
integrals of flat-space string theory. We derived
all-multiplicity monodromy relations for the open-string
building blocks and a KLT factorisation for their
closed-string counterparts, extending the four-point
AdS constructions of~\cite{Alday:2025bjp,Alday:2025cxr}.

Beyond four points, the worldsheet integrals depend
on several independent puncture positions. The main challenge, however, lies in analytically
continuing the polylogarithmic insertions between
ordering chambers. This requires comparing distinct
tangential base-point normalisations through
non-commuting transport factors associated with
successive puncture collisions. We showed that the
auxiliary regulators cancel in the local crossing
factors, which can then be combined in the order
prescribed by the continuation path. The same
chamber-dependent transport enters the KLT
factorisation, where the single-valued pairing
requires appropriately transported right-moving
building blocks. The resulting KLT kernel retains an iterative
structure, with an essential ordering of
non-commuting crossing factors.
Both relations reproduce the known four-point AdS
results and reduce to their flat-space counterparts
when $e_{ij}\to0$. Their higher-point structure is
illustrated explicitly at five points in
Appendix~\ref{app:five-point-example}.

The structure uncovered here suggests several future directions. First, it would be useful to understand more directly the space of functions generated by our integrals. At fixed multiplicity, the polylogarithmic insertions belong to a finite-dimensional space at each weight. This follows from Brown's fibration-basis for polylogarithms on $\mathcal M_{0,n}$, where the relevant algebra is organised as a tensor product of free shuffle algebras on $(2,3,\ldots,n-2)$ generators \cite{brown2006multiplezetavaluesperiods}. In our conventions, this is precisely the structure of the generating series $G_{\mathbb I}$, whose successive one-variable factors have alphabets of sizes $(n-2,n-3,\ldots,2)$. Thus, the multivariable nature of the problem does not lead to an uncontrolled functional space. For instance, at five points, the number of basis elements at weights $(0,1,2,3,4)$ is $(1,\ 5,\ 19,\ 65,\ 211)$. In concrete bootstrap applications, this raw functional space should be further reduced by physical constraints such as crossing symmetry, regularity, and OPE limits. At four points, the monodromy relations satisfied by the colour-ordered AdS Veneziano amplitudes drastically constrain the allowed MPL ansatz for the curvature expansion \cite{Alday:2025cxr}. 

A related question concerns the transcendental weights
that appear at each order in the curvature expansion.
The weight-counting argument of
Section~\ref{sec:worldsheet-motivation} is consistent
with the maximal transcendentality patterns observed
in flat-space and AdS string amplitudes
\cite{Stieberger:2013wea,Stieberger:2014hba,Brown:2019wna,Alday:2023mvu}. In the open-string AdS Veneziano case, the single-valued
structure of the weight-three MPL ansatz on the real line
provides further constraints on higher curvature corrections
\cite{Alday:2024ksp}.
Our construction may provide a natural framework for
extending these ideas to higher-point holographic correlators,
including the recently obtained five-point stringy correction
to the $20'$ correlator in $\mathcal N=4$ SYM
\cite{VilasBoas:2025vvw}.

Another natural direction concerns the inverse of the AdS KLT kernel. In flat space, the inverse string KLT kernel admits a geometric interpretation in terms of twisted intersections of integration cycles and reduces, in the field-theory limit, to the bi-adjoint double-partial amplitude matrix \cite{Mizera:2016jhj,Mizera:2017cqs}. The four-point AdS answer suggests a non-commutative version of this story \cite{Kakkad:2025klm}. At higher multiplicity, one may expect analogous local contributions associated with boundary divisors of $\mathcal M_{0,n}$. It would be interesting to establish a direct connection between this geometric picture and our contour construction.  Determining which combinations of these building
blocks arise in physical AdS string amplitudes
remains an important open problem.

\paragraph{Note added.}
After the first version of this paper appeared, an
intersection-theoretic construction of the inverse AdS KLT
kernel was developed in~\cite{Tao:2026qeh}, providing further
geometric support for the all-multiplicity KLT structure
derived here.

\paragraph{Acknowledgements.}
We thank Oliver Schlotterer for many fruitful discussions and for fostering the exchange of ideas that led to this collaboration. We are also grateful to Fernando Alday and Shijie Zhang for useful conversations. A.S. would further like to thank Lance Dixon for valuable insights. The work of R.S.P. is supported by the São Paulo Research Foundation (FAPESP) through grant 2022/05236-1. The work of Y.T. is supported by the National Natural Science Foundation of China (NSFC) (Grant No.~124B2094). 

\appendix
\section{Five-point example}
\label{app:five-point-example}

\begin{figure}[htbp]
  \centering
  \includegraphics[width=0.65\linewidth]{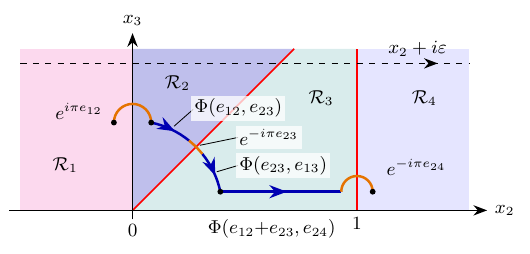}
  \caption{Five-point contour (dashed) and associator transport (blue);
  orange arcs mark the crossings. The displayed $e_{ij}$ phases
  combine with the Koba--Nielsen phases to give $E_{ij}$.}
  \label{fig:five-point-transport}
\end{figure}

At five points, the polylogarithmic insertion depends on two
moduli, and its continuation between ordering chambers
involves non-trivial KZ transport.
For fixed $0<x_3<1$, the $x_2$ contour passes through
the four chambers $(21345)$, $(12345)$, $(13245)$,
and $(13425)$, as shown in Figure~\ref{fig:five-point-transport}.
Starting from the canonical chamber $(12345)$,
normalised at $0<x_2\ll x_3\ll 1$, the continuation
produces ordered products of Drinfeld associators
that enter both the monodromy and KLT relations.

\paragraph{Monodromy.}
The leftward crossing at $x_2=0$ gives $e^{i\pi E_{12}}$.
For the two rightward crossings, at $x_2=x_3$ and $x_2=1$,
the factors accompanying the canonical ordered integrals are
\begin{align}
  C_{23}&=\Phi(e_{23},e_{13})\,e^{-i\pi E_{23}}\,\Phi(e_{12},e_{23}),
  \label{eq:five-point-C23}\\
  C_{24}&=\Phi(e_{24},e_{14}+e_{34})\,e^{-i\pi E_{24}}\,
  \Phi(e_{12}+e_{23},e_{24}).
  \label{eq:five-point-C24}
\end{align}
In $C_{23}$, the rightmost associator transports to the collision,
while the leftmost restores the canonical normalisation in $(13245)$.
In $C_{24}$, the punctures at $0$ and $x_3$ act as a cluster, hence
the effective residue $e_{12}+e_{23}$. Its leftmost associator, not
shown in the figure, restores the canonical gauge and tangential
base-point normalisation of $(13425)$.
With $Z_{\tau}=Z^{(\mathrm{AdS})}(\tau\mid\rho)$, the contour relation is
\begin{equation}
 Z_{21345}e^{i\pi E_{12}}+Z_{12345}
 +Z_{13245}C_{23}+Z_{13425}C_{24}C_{23}=0.
 \label{eq:five-point-monodromy}
\end{equation}
The order $C_{24}C_{23}$ matters. In the flat-space limit,
$e_{ij}\to0$, the factors reduce to $C_{23}\to e^{-i\pi s_{23}}$
and $C_{24}\to e^{-i\pi s_{24}}$, giving the standard monodromy relation.

\paragraph{KLT.}
The same ordered transport enters the closed-string factorisation.
In the bases $\alpha,\beta\in\{(23),(32)\}$, we have
\begin{equation}
 J^{(\mathrm{AdS})}(\tau\mid\rho)
 =\begin{pmatrix}Z_{23145}&Z_{32145}\end{pmatrix}
 \begin{pmatrix}S[23|23]&S[23|32]\\ S[32|23]&S[32|32]\end{pmatrix}
 \begin{pmatrix}\widetilde Z_{12345}\\ \widetilde Z_{13245}\end{pmatrix},
 \label{eq:five-point-KLT}
\end{equation}
where $S=S_{(\mathrm{AdS})}$ and the right-moving blocks
$\widetilde Z$ are as defined in
\eqref{eq:right-moving-Z-def}. For example, the first
column of the kernel reads
\begin{align}
 S[23|23]
 &=\frac{1}{\pi^2}A\sin(\pi E_{13})B\sin(\pi E_{12}),\\
 S[32|23]
 &=\frac{1}{2i\pi^2}
 \left(e^{i\pi E_{23}}A e^{i\pi E_{13}}
      -e^{-i\pi E_{23}}A e^{-i\pi E_{13}}\right)
      B\sin(\pi E_{12}),
 \label{eq:five-point-KLT-column}
\end{align}
with $A=\Phi(e_{13},e_{23})$ and $B=\Phi(e_{12},e_{13})$.
The second column follows from the reversed contour-deformation order.
In the flat-space limit $e_{ij}\to0$, the associators
$A,B\to1$, and the displayed entries reduce to
\begin{align*}
S[23|23]
&\longrightarrow
\frac{\sin(\pi s_{13})\sin(\pi s_{12})}{\pi^2},\\
S[32|23]
&\longrightarrow
\frac{\sin(\pi(s_{13}+s_{23}))\sin(\pi s_{12})}{\pi^2}.
\end{align*}
The $2\times2$ KLT structure is already present in flat space;
the AdS deformation lies in the ordered, non-commutative
associator transport entering its matrix elements.

\bibliographystyle{JHEP}
\bibliography{references}

\end{document}